\documentclass{article}
\usepackage[whole]{bxcjkjatype}
\usepackage[unicode]{hyperref}
\usepackage[left=2.50cm, right=2.50cm, top=2.50cm, bottom=2.50cm]{geometry} 
\usepackage{helvet}
\usepackage{amsmath, amsfonts, amssymb} 
\usepackage{amsmath} 
\usepackage[english]{babel}
\usepackage{graphicx}   
\usepackage{subfig}
\usepackage{subfloat}
\usepackage{tikz}

\usepackage{makecell} 
\usepackage{url}        
\usepackage{bm}         
\usepackage{multirow}
\usepackage{verbatim}   
\usepackage{cite}
\usepackage{booktabs}
\usepackage{algorithm}
\usepackage{algorithmic}

\usepackage{fancyhdr} 
\newcommand{\setParDis}{\setlength {\parskip} {1.5cm} }
\newcommand{\setParDef}{\setlength {\parskip} {0pt} }

\usepackage{multicol}

\usepackage[T1]{fontenc}
\usepackage[utf8]{inputenc}
\usepackage{authblk}
\usepackage[misc]{ifsym} 

\usepackage{stfloats}

\title{\textbf{House Price Valuation Model Based on Geographically Neural Network Weighted Regression: The Case Study of Shenzhen, China}}
\author[a]{Zimo WANG}
\author[a]{Yicheng WANG}
\author[a,b]{\textrm{\Letter} Sensen WU\thanks{CONTACT: Wang Zimo wangzimo@zju.edu.cn, Wang Yicheng wangyicheng@zju.edu.cn, Wu Sensen wusensengis@zju.edu.cn}}
\affil[a]{\small School of Earth Sciences, Zhejiang University, China}
\affil[b]{\small Zhejiang Key Laboratory of Resources and Environmental Information System, Hangzhou, China}

\date{(Dated: \today)}

\begin{document}
	\maketitle
	\noindent{\bf Abstract: }Confronted with the spatial heterogeneity of real estate market, some traditional research utilized Geographically Weighted Regression (GWR) to estimate the house price. However, its kernel function is non-linear, elusive, and complex to opt bandwidth, the predictive power could also be improved. Consequently, a novel technique, Geographical Neural Network Weighted Regression (GNNWR), has been applied to improve the accuracy of real estate appraisal with the help of neural networks. Based on Shenzhen house price dataset, this work conspicuously captures the weight distribution of different variants at Shenzhen real estate market, which GWR is difficult to materialize. Moreover, we focus on the performance of GNNWR, verify its robustness and superiority, refine the experiment process with 10-fold cross-validation, extend its application area from natural to socioeconomic geospatial data. It's a practical and trenchant way to assess house price, and we demonstrate the effectiveness of GNNWR on a complex socioeconomic dataset.\\
	
	\begin{multicols}{2}
		\section{Introduction}
		As one of the countries with the fastest urbanization process, China has steadily rising housing prices in the past decades, especially in its major cities. Affected by the COVID-19 in 2020, the world's major economies entered a liquidity easing cycle, and housing prices in many cities in China rose significantly.\cite{ref1} On this basis, several Chinese cities, such as Shenzhen, Xi'an, Chengdu, have implemented second-hand housing transaction reference price, which is used to curb housing price rising. The reference price provides us with a reasonable valuation for the housing prices with slightly bubbles.
		
		Housing price is closely related to the life of new urban residents, and it is also an economic index that the government needs to pay close attention to. Exploring the spatial distribution pattern of housing price has great practical significance and guiding value for government regulation, individual house purchase or third-party valuation.
		
		To model and estimate house prices, different models have been developed by many scholars. In 1972, Rosen proposed the Hedonic model, which aims to measure property prices using a number of environmental factors. The early studies mainly consisted of three components: locational traits, structural traits, and neighborhood traits, i.e., residential prices are mainly a function of these three characteristics and are approximately linearly related in an exponentially corrected manner.\cite{ref2,ref3} A number of subsequent studies have demonstrated the relative validity of this model, and measures of these factors are able to estimate more accurately the positive or negative correlation between each independent variable and house prices. For example, HENRY M.K. MOK's modeling of house prices in Hong Kong in 1994 showed that house prices were significantly negatively correlated with the age of the house, distance from the CBD, and significantly positively correlated with the floor.\cite{ref4} As times progressed, more and more independent variables were taken into account and more statistical indicators were added to test the validity of the model. Further studies have partially incorporated land use planning, and also accessibility in terms of transportation.\cite{ref5} In recent years, related residential house price studies have incorporated a variety of external environmental factors such as natural landscape and neighborhood size to analyze the impact on house prices.\cite{ref6} However, these models have constantly encountered problems in dealing with spatial heterogeneity and spatial non-stationarity, i.e., the same independent variable has different effects on house prices in different regions. Ordinary Hedonic models can only model a certain independent variable with constant coefficients, but the real situation often influenced by spatial factors. For example, in suburban areas, transportation conditions dominate house prices and the quality of nearby schools does not matter. However, in downtown areas, the quality of schools near homes might be more critical and nearby transportation conditions are relatively less important. This is something that cannot be analyzed by ordinary Hedonic models.
		
		Further, taking into account the spatial heterogeneity of the different influencing factors, Geographically Weighted Regression (GWR) methods are proposed that allow the coefficients to change at different locations.\cite{ref7,ref8} The method can be understood as a local weighted linear regression for each local area, and the weights fully take into account the effects of adjacent data points according to the first law of geography proposed by Tobler.\cite{ref9} In order to build a more satisfying model for the relationships between the house price and the area, Brunsdon and Fotheringham have mentioned several key questions GWR faced: the selection of the variables, bandwidth, and the spatial autocorrelation of error after proposing GWR. \cite{ref10} Many scholars have made attempts on this basis. For example, in 2011 Jijin Geng et al. had used the GWR model to model house prices in Shenzhen. Compared with the Ordinary Linear Regression (OLR) model, the R square improved from 0.56 to 0.79.\cite{ref11} Zhang et al. used mixed geographically weighted regression to model the rent in Nanjing, i.e., some variables were locally weighted according to geographical location, while some variables were globally weighted, and good results were achieved.\cite{ref12} Binbin Lu added non-Euclidean distance to GWR, and for some geographic elements that do not obey the standard linear measure, this model achieves better results on the spatial proximity measure of London and can have better estimation performance for house prices.\cite{ref13,ref14}
		
		However, the ability of GWR to express nonlinear spatial relationships is quite limited. Therefore, many scholars have resorted to artificial intelligence methods, which have developed rapidly in recent years, to model house prices using their superb fitting ability to nonlinear relationships.\cite{ref15,ref16} Although the estimation performance of neural network models is usually superior to that of GWR models, the spatial distributions obtained by these models are not entirely reasonable and the constructed regression relationships are difficult to interpret spatially, because they ignore the spatial properties of residential price regression relationships.
		
		In recent years, based on the idea of geographic weighting of GWR, Sensen Wu proposed a Geographically Neural Network Weighted Regression (GNNWR) model by combining OLR and neural network models.\cite{ref17,ref18} Based on the powerful learning ability of neural networks, the potential spatial non-stationarity and complex nonlinear features in regression relations can be well handled. In the current study, GNNWR has effectively solved many problems and has performed well in modeling the ecological environment of nearshore seas\cite{ref19}, also showed superior explanatory power in the estimation of spatial PM 2.5 concentrations in China.\cite{ref20}
		
		On February 8, 2021, the Shenzhen Real Estate and Urban Construction Development Research Center released reference prices for second-hand housing transactions for the city's 3,595 residential quarters.\cite{ref21} Based on this dataset, a residential price valuation model can be developed, which covers various factors such as property endogenous variables, subway, and school district conditions. This study attempts to build a residential price valuation model with the help of a relatively new tool, GNNWR, in an attempt to deal with the spatial heterogeneity and spatial non-stationarity present in this data.\cite{ref18}
		
		In summary, this study aims to put the GNNWR model into practice in the socioeconomic field, establish a residential price valuation model based on the reference price data of second-hand housing transactions in Shenzhen, realize the accurate fitting of the spatial heterogeneity and nonlinear relationship of multiple environmental factors in the modeling, and then obtain a more accurate house pricing model than GWR method, with the spatial distribution of multiple factors' influence coefficients. It can provide reference for residential valuation, land auctions, and the reference prices of second-hand housing transactions in other cities.

		\section{Study Area, Data Sources and Research Methods}
		\subsection{Shenzhen House Price Profile}
		
		In 1980, Shenzhen Special Economic Zone was established. It is adjacent to Hong Kong in the south, located in the west of the Pearl River Estuary in Guangdong Province, China. With its geographical advantages close to Hong Kong and the policy support, Shenzhen has now become the third largest sorted by GDP, with nine districts under its jurisdiction. According to the announcement data of the Seventh National Census, the population of Shenzhen has reached 17.56 million. Even under the impact of COVID-19, Shenzhen's regional GDP reached 2767.024 billion RMB in 2020, an increase of 3.1\% over 2019.\cite{ref23}
		
		Owing to the increasement of population, with great economic conditions and perfect business environment, the house prices in Shenzhen are also rising. In the short term, affected by the loose liquidity stimulated spurred by the COVID-19, Shenzhen real estate market in 2020 was quite prosperous. The investment in real estate development increased by 16.4\% over the previous year; the residential construction area increased by 21\%; and the sales area of commercial housing increased by 17.3\%, which led to the further rise of house prices as a whole. According to the second-hand housing data of 70 large and medium-sized cities released by the National Bureau of Statistics of China in 2020, Shenzhen's real estate market rose by 14.1\% throughout the year. As the only city with an increase of more than 10\% in China, it ranked first in the growth rate.
		
		In order to suppress the excessive growth of house prices, in  February 2021, Shenzhen Real Estate and Urban Construction Development Research Center formed the reference price of second-hand housing transactions in 3,595 residential quarters, based on the government recorded transaction prices of second-hand housing and the surrounding first-hand housing price.\cite{ref24}
		
		From the perspective of data modeling, the Shenzhen data were selected for the study mainly due to the following factors. First, the reference price of second-hand housing transactions has itself undergone considerable evaluation compared to other data. It averages out the differences in different house types and floors, and also combines government recorded transaction prices and surrounding first-hand housing prices, removing short-term heat and bubbles and reflecting a relatively accurate valuation result for a property. Secondly, Shenzhen's urban development is more natural. There is no important political center, relics or slums affecting urban planning. Finally, the reference prices are introduced in a uniform batch, with a large amount of data and influence. Modeling of the reference price of second-hand house transactions in Shenzhen can provide reference for more cities to introduce similar measures.
		
		\subsection{Experimental Data}
		
		\begin{figure}[H]
			\centering
			\includegraphics[width=0.45\textwidth]{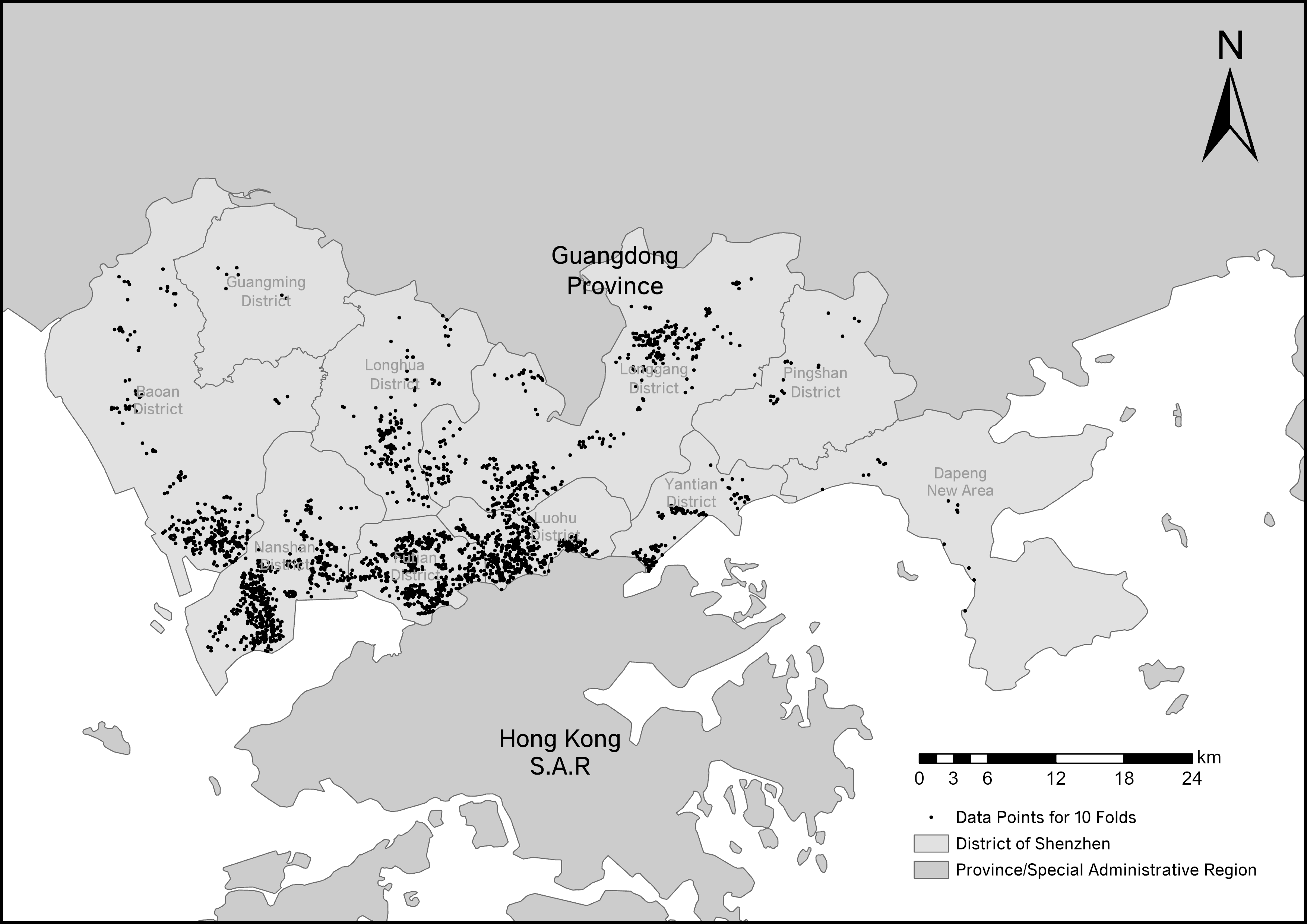}
			\caption{Datapoints for 10 Folds}
		\end{figure}
	
		\begin{figure}[H]
			\centering
		 	\includegraphics[width=0.45\textwidth]{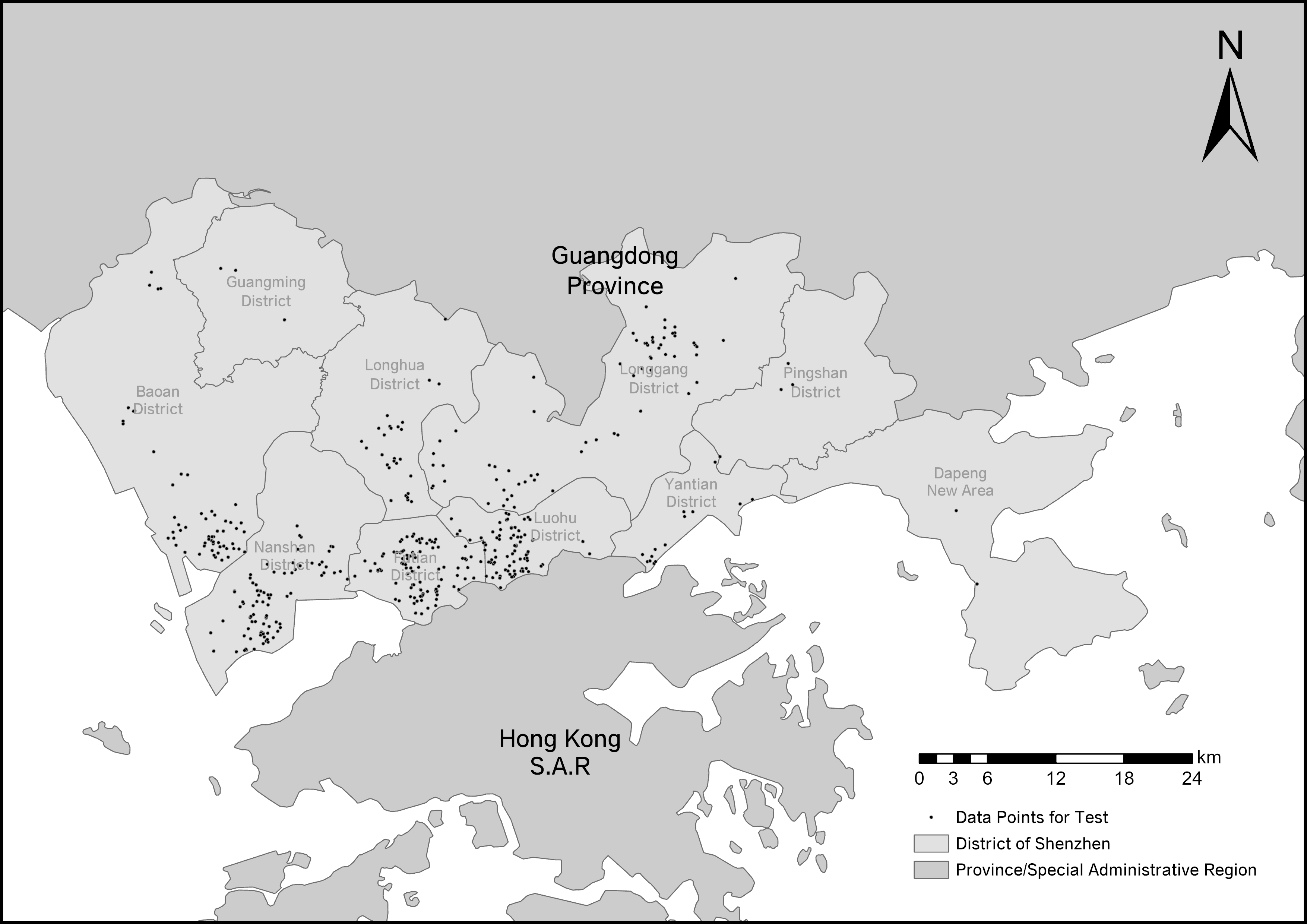}
		 	\caption{Datapoints for Test Set}
		\end{figure}

		The total number of complete and effective initial data records obtained in this study is 2871, covering 2871 residential quarters in Shenzhen. The specific data source is \url{https://shenzhen.qfang.com/}. In this study, it is divided into the following three categories according to its functions:
		
		\begin{enumerate}
			\item Latitude and longitude of the property: The data crawled by this property are latitude between 22.484310 N and 22.788011 N and longitude between 113.814605 E and 114.498340 E. The latitude and longitude coordinates used for GNNWR were converted to WGS 1984 50N coordinate system after projection conversion.
			
			\item Endogenous Variable of the property: age of building (AB), number of parking spots (NPS), management fee (MF), number of buildings (NB), green ratio (GR), plot ratio (PR). The house age is calculated according to 2021 minus the construction age. If the construction age is a rage, the completion time is taken. If the property fee is a range, it shall be calculated according to the average value of the upper and lower bounds. Greening rate and plot ratio are defined as follows: 
			$$GR=\dfrac{S_{Vegetation}}{S_{Total}}$$
			$$PR=\frac{S_{Total\ Floor}}{S_{Land\ Area}}$$
			
			\item Environment related variables: sea distance (SD), quality of available public schools (QAPS), number of subway stations within 1 km radius (NSS), distance to the nearest subway station (DSS). Where SD is calculated by taking the position to the nearest coast nearby, with DSS units in meters. QAPS is calculated as following method: we divide schools into 4 types, including provincial key schools, city key schools, district key schools and ordinary schools. We give different points for different schools: 1 point for ordinary junior high school, 2 points for district key junior high school, 3 points for city key junior high school, 4 points for provincial key junior high school, 1.5 points for ordinary elementary school, 2.5 points for district key elementary school, 3.5 points for city key elementary school, and 4.5 points for provincial key elementary school. Finally, the points of the best school in a school district will be taken as QASP.
			
		\end{enumerate}
		
		\subsection{Research Methodology}
		
		\noindent{\bf Geographically Weighted Regression (GWR): }In the classic ordinary linear regression (OLR) model, dependent variable and independent variables can be expressed by the regression equation:
		$$y_{i}=\beta_{0}+\sum_{k=1}^{k}\beta_{k}x_{ik}+\epsilon_{i} i=1,2,3\dots$$
		where $\beta_{0}$ is regression constant; $\beta_{1},\dots,\beta_{p}$ are regression coefficients; $\epsilon_{i}$ is the error term of the sample with the mean value zero and constant variance $\sigma^2$.
		
		Moreover, its coefficient can be estimated as:
		$$\bm{\widehat{\beta}}=\bm{(X^TX)^{-1}X^Ty}$$
		where
		$$
		\bm{X}=
		\begin{bmatrix}
			1 &x_{11} &x_{12} &\cdots &x_{1p} \\
			1 &x_{21} &x_{22} &\cdots &x_{2p} \\
			\vdots &\vdots &\vdots &\ddots &\vdots \\
			1 &x_{n1} &x_{n2} &\cdots &x_{np} 
		\end{bmatrix}, 
		$$
		$$
		\bm{y}=
		\begin{bmatrix}
			y_{1} \\
			y_{2} \\
			\vdots \\
			y_{n}
		\end{bmatrix},  
		\bm{\widehat{\beta}}=
		\begin{bmatrix}
			\beta_{0} \\
			\beta_{1} \\
			\vdots    \\
			\beta_{p} 
		\end{bmatrix}
		$$
		
		In fact, the regression coefficient calculated by OLR model is the best unbiased estimation of all sample points, which can be regarded as the average relationship in the whole study area. The spatial variation of this relationship can be regarded as different fluctuations of the "average relationship" caused by spatial non-stationarity.
		
		Based on the first law of geography, some scholars proposed a spatial weighted regression (GWR) model, trying to change the regression coefficient from global to local, and change the weight of adjacent points according to different distances in the regression framework. GWR model defines spatial non-stationarity as\cite{ref7,ref26}: 	
		$$y_{i}=w_{0}(u_{i},v_{i})\times\beta_{0}+\sum_{k=1}^{p}w_{k}(u_{i},v_{i})\times\beta_{k}x_{ik}+\epsilon_{i}$$		
		where $\beta_{0}(u_{i},v_{i})\times\beta_{0}$; $\beta_{k}(u_{i},v_{i})\times\beta_{k}$. Therefore, we can regard $w_{0}(u_{i},v_{i})$ as the non-stationarity weight of the regression constant $\beta_{0}$, and $w_{k}(u_{i},v_{i})$ represents the non-stationarity weight of regression coefficient $\beta_{k}$. Substituting the estimated value of OLR $\widehat{\beta}_{k}$ into the above formula, the estimated value  can be obtained as follows: 
		$$\widehat{y}_{i}=\sum_{k=0}^{p}\widehat{\beta}_{k}(u_{i},v_{i})x_{ik}
		$$
		
		The estimator in matrix form can be expressed as:
		$$\widehat{y}_{i}=\bm{x^{T}_{i}}(\bm{X^{T}W}(u_{i},v_{i})\bm{X})^{-1}\bm{X^TW}(u_{i}, v_{i})\bm{y}$$
		The Spatial weight matrix $\bm{W}(u_{i}, v_{i})$ can be expressed as:
		$$
		\bm{W}(u_{i},v_{i})\triangleq
		\begin{bmatrix}
			w_{1}(u_{i}, v_{i}) &0 &\cdots &0 \\
			0 &w_{2}(u_{i}, v_{i}) &\cdots &0 \\
			\vdots &\vdots &\ddots &\vdots \\
			0 &0 &\cdots &w_{n}(u_{i}, v_{i})
		\end{bmatrix}
		$$
		
		In GWR model, the weight kernels usually use Gaussian, bi-square, tri-cube and exponential functions. These functions can relatively simply express the complex relationship between spatial proximity (i.e. spatial distance) and spatial non-stationarity (i.e. spatial weight).
		
		It should be noted that there are different ways to select the function in the spatial weight matrix. Different selection methods directly affect the final modeling accuracy.
		
		The Gaussian weighted function can be expressed as:
		$$w_{ij}=e^{-\dfrac{d_{ij}^s}{b^2}}$$
		
		where $d_{ij}^s$ is the distance between points i and j; b, the bandwidth, producing a declining effect relative to $d_{ij}^s$, has different methods to select: for the fixed Gaussian weight function, the bandwidth is the same at each point and is a constant in the same model; for the adaptive Gaussian weight function, the bandwidth is different at each point, and the point distance closest to the point is often taken as the value of bandwidth. In any case, the Gaussian weight function requires a variable input, that is, the distance range (fixed bandwidth) or the number of major adjacent features (dynamic bandwidth).
		
		The bi-square weighted function can be expressed as:
		
		$$
		w_{ij}=\begin{cases}
			[1-(d^s_{ij}/b_i)^2]^2, &d_{ij}^s<b_{i}; \\
			0, &\text{the others}.
			\end{cases}
		$$
		
		where $d^{s}_{ij}$ is the distance between two points; $b_{i}$ is the bandwidth. It is also divided into fixed type and adaptive type according to the above method.
		
		This model is built using adaptive functions, i.e., an input variable is needed to select the number of major neighboring elements, and the AICc criterion is used to determine whether it is more preferable.\cite{ref27}
		
		\setParDis
		
		\noindent{\bf Geographically Neural Network Weighted Regression (GNNWR): }Similarly, based on the non-stationarity in the spatial relationship, GNNWR goes further than GWR, trying to further accurately describe the fluctuation level of spatial non-stationarity on the regression relationship at different locations. The key step of GWR is the selection and construction of spatial weight matrix function. On this basis, GNNWR attempts to go further and find an appropriate spatial weight matrix function with the help of neural network.

		\setParDef
		
		To accurately fit the complex relationship between spatial distance and spatial weight, GNNWR designs a spatial weighted neural network (SWNN) to achieve the neural network expression of weight kernel function. Specifically, SWNN takes the spatial distance between points as the input layer and the spatial weight matrix as the output layer, and selects the appropriate number of hidden layers according to the modeling needs. The spatial weight calculation of the points corresponds with:
		$$\widehat{y_i}=\bm{x_i^T\widehat{\beta}}(u_{i}, v_{i})=\bm{x_i^TW}(u_i, v_i)(\bm{X^TX})^{-1}\bm{X^Ty}$$
		
		where $\bm{W}(u_i, v_i)$ is the spatial weight matrix as:
		$$
		\bm{W}(u_{i},v_{i})\triangleq
		\begin{bmatrix}
			w_{0}(u_{i}, v_{i}) &0 &\cdots &0 \\
			0 &w_{1}(u_{i}, v_{i}) &\cdots &0 \\
			\vdots &\vdots &\ddots &\vdots \\
			0 &0 &\cdots &w_{p}(u_{i}, v_{i})
		\end{bmatrix}
		$$
		
		That is, this matrix is the result of function $\bm{W}:\mathsf{R}^2\rightarrow\mathsf{R}^{(1+p)\times(1+p)}$. SWNN further considers the existence of an intermediate variable $[d_{i1}, d_{i2}, d_{i3}, \cdots, d_{in}]$ and matric $\bm{W}(u_i, v_i)$ is a function of variable $[d_{i1}, d_{i2}, d_{i3}, \cdots, d_{in}]$, where $d_{ij}$ is the distance from point i to sample point j. Thus, the GNNWR-based house price estimation model is shown as Figure \ref{NN structure}: 
		
		\begin{figure*}[htbp]
			\centering
			\includegraphics[width=0.9\textwidth]{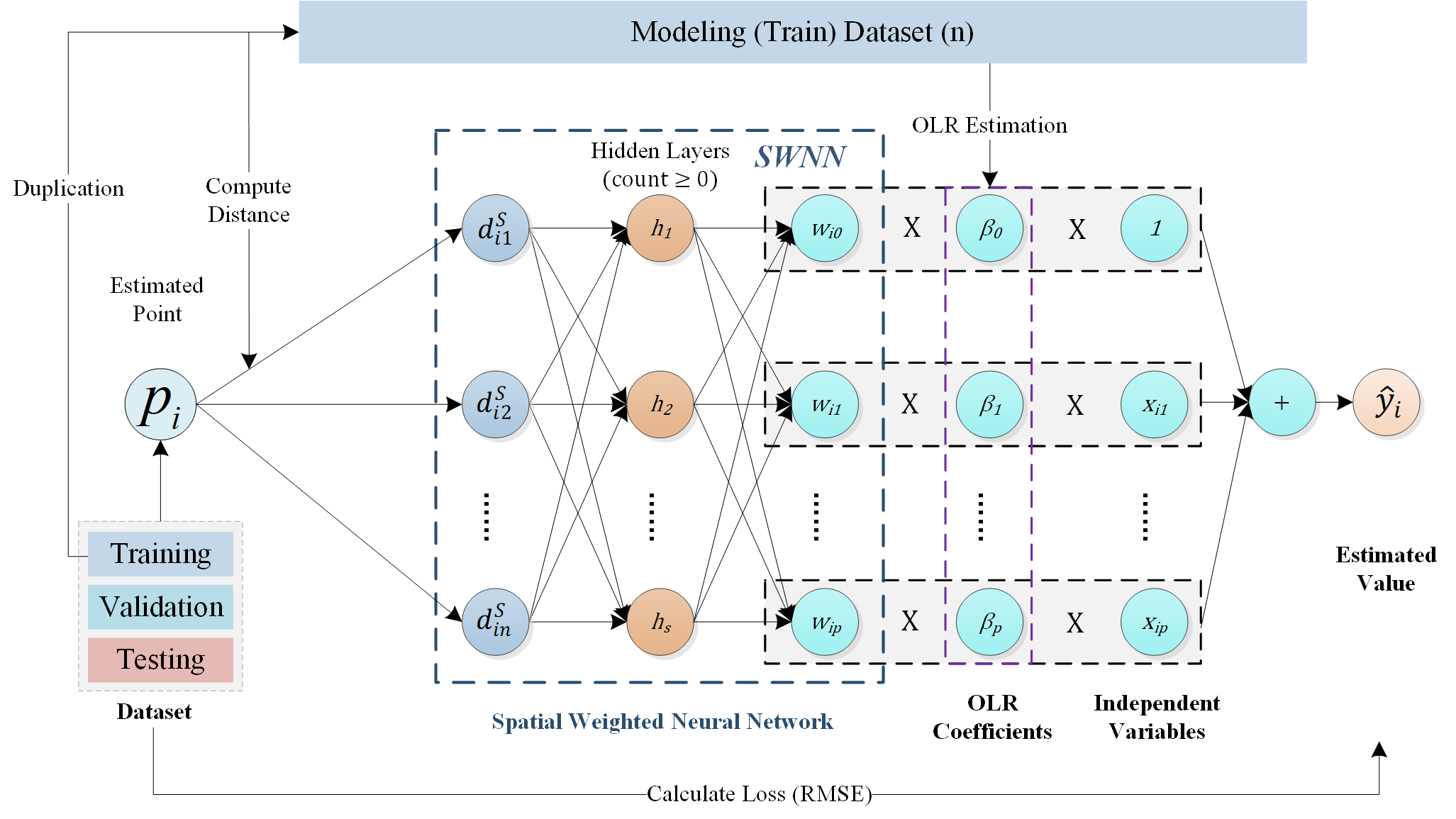}
			\caption{Network Structure for GNNWR}
			\label{NN structure}
		\end{figure*}

		\subsection{Research Indicators}
		

		\noindent{\bf Significance Test Statistics for Spatial Non-stationarity: }To test whether the relationship has significant spatial non-stationarity, we use the residual sum of squares and its approximated distribution deduced by Leung i.e.\cite{ref28} and Wu\cite{ref17}, for significance tests of GNNWR and GWR modeling results.
		
		Firstly, express the hat matrix of GNNWR as:
		$$
		S_{GNNWR}\triangleq
		\begin{bmatrix}
			x_1^TW(u_1, v_1)(X^TX)^{-1}X^T \\
			x_2^TW(u_2, v_2)(X^TX)^{-1}X^T \\
			\vdots \\
			x_n^TW(u_n, v_n)(X^TX)^{-1}X^T
		\end{bmatrix}
		$$
		$$\delta_i\triangleq tr\{[(I-S)^T(I-S)]^i\},i=1,2\cdots$$
		
		The statistical quantities $F_1$ is obtained as:
		$$F_1=\dfrac{RSS_{GNNWR}/\delta_1}{RSS_{OLR}/(n-p-1)}$$
		
		The distribution of $F1$ can be approximated as F distribution，where $\dfrac{\delta_1^2}{\delta_2}$ is the degree of freedom of the numerator and $n-p-1$ is the degree of freedom of the denominator. That is, given a significance level $\alpha$, if the inequality $F_1<F_{1-\alpha}(\frac{\delta_1^2}{\delta_2},n-p-1)$ holds, it can be determined that the regression relationship has significant spatial non-smoothness, otherwise the spatial non-smoothness is not significant.
		
		Second, the significance of the spatial non-stationarity can also be checked for each independent variable one by one. The null hypothesis here is that the weight of this independent variable is the same everywhere in the space. The alternative hypothesis is that the weight of this independent variable differs in at least one place in each part of the space. First, define  the variance of the weight of the kth independent variable over the n data points.
		
		$$V^2_k\triangleq\frac{1}{n}\sum_{i=1}^{n}(\widehat{\beta}_{ik}-\frac{1}{n}\sum_{i=1}^{n}\widehat{\beta}_{ik})^2$$
		
		Define $e_k$ as a n-rank vector with the $(k+1)^{th}$ element having value 1 and other having value 0. Define  as a square matrix of order n with each element having value 1.
		
		$$B_k\triangleq
		\begin{bmatrix}
			e_k^TW(u_1,v_1)(X^TX)^{-1}X^T \\
			e_k^TW(u_2,v_2)(X^TX)^{-1}X^T \\
			\vdots \\
			e_k^TW(u_n,v_n)(X^TX)^{-1}X^T
		\end{bmatrix}
		$$
		
		$$\gamma_{ik}\triangleq tr\{[\frac{1}{n}B_k^T(I-\frac{1}{n}J)B_k]^i\},i=1,2\cdots$$
		
		The statistical quantities $F_2$ is obtained as:
		$$F_2(k)=\frac{V_k^2/\gamma_{1k}}{\widehat{\sigma}^2}$$
		
		The distribution of $F_2(k)$ can be approximated as F distribution, where $\widehat{\sigma}^2$ is the mean square error, $\frac{\gamma^2_{1k}}{\gamma_{2k}}$ is the degree of freedom of the numerator and $\frac{\delta_1^2}{\delta_2}$ is the degree of freedom of the denominator. That is, given a significance level $\alpha$, if the inequality $F_2(k)>F_{\alpha}(\frac{\gamma^2_{1k}}{\gamma_{2k}},\frac{\delta_1^2}{\delta_2})$ holds, the null hypothesis can be rejected and the variable k is determined to have significant spatial non-stationarity, otherwise the spatial non-stationarity is not significant.
		
		\setParDis

		\noindent{\bf Indicators of Model Performance: }The paper uses the following metrics to evaluate the performance of the model. First in the AICc guidelines, the correction of Akaike information criteria ($AIC_c$)\cite{ref7} is as follows:

		\setParDef
		$$AIC_c=nln(\widehat{\sigma}^2)+nln(2\pi)+n\frac{n+tr(S)}{n-2-tr(S)}$$
		
		The method is applicable for both GWR and GNNWR. In practice, the smaller the value, the better\cite{ref27}, and we use $AIC_c$ to select the appropriate input parameters for the GWR model. Other measures of model performance include: coefficient of determination ($R^2$), root mean square error (RMSE), mean absolute error (MAE), and mean absolute percentage error (MAPE). The definitions are as follows: 
		$$R^2=1-\frac{\sum_{i=1}^{n}(y_i-\widehat{y}_i)^2}{\sum_{i=1}^{n}(y_i-\overline{y}_i)^2}$$
		
		$$RMSE=\sqrt{\frac{\sum_{i=1}^{n}(y_i-\widehat{y}_i)^2}{n}}$$
		
		$$MAE=\frac{\sum_{i=1}^{n}|y_i-\widehat{y}_i|}{n}$$
		
		$$MAPE=\frac{1}{n}\sum_{i=1}^{n}|\frac{y_i-\widehat{y}_i}{y_i}|\times100\%$$
		
		Among them, $\overline{y}$ is the average of the observed values; $\widehat{\sigma}^2$ is the mean square error of the model and p is the effective degree of freedom of the model.

		\subsection{Neural Network Design and Implementation}
		The model uses a traditional neural network and the specific process is shown in Figure 4. Combined with 10-fold cross-validation can ensure the robustness and reliability of the algorithm. All layers of the spatially weighted neural network are full-connected with each other, and the Dropout technique proposed by Srivastava et al. is applied to improve the generalization ability of the model.\cite{ref32} Each hidden layer is combined with the Batch Normalization technique. The parameter initialization is adopted by the method proposed by He\cite{ref33} et al. and the activation function is adopted by PReLU.
		
		\begin{figure}[H]
			\centering
			\includegraphics[width=0.45\textwidth]{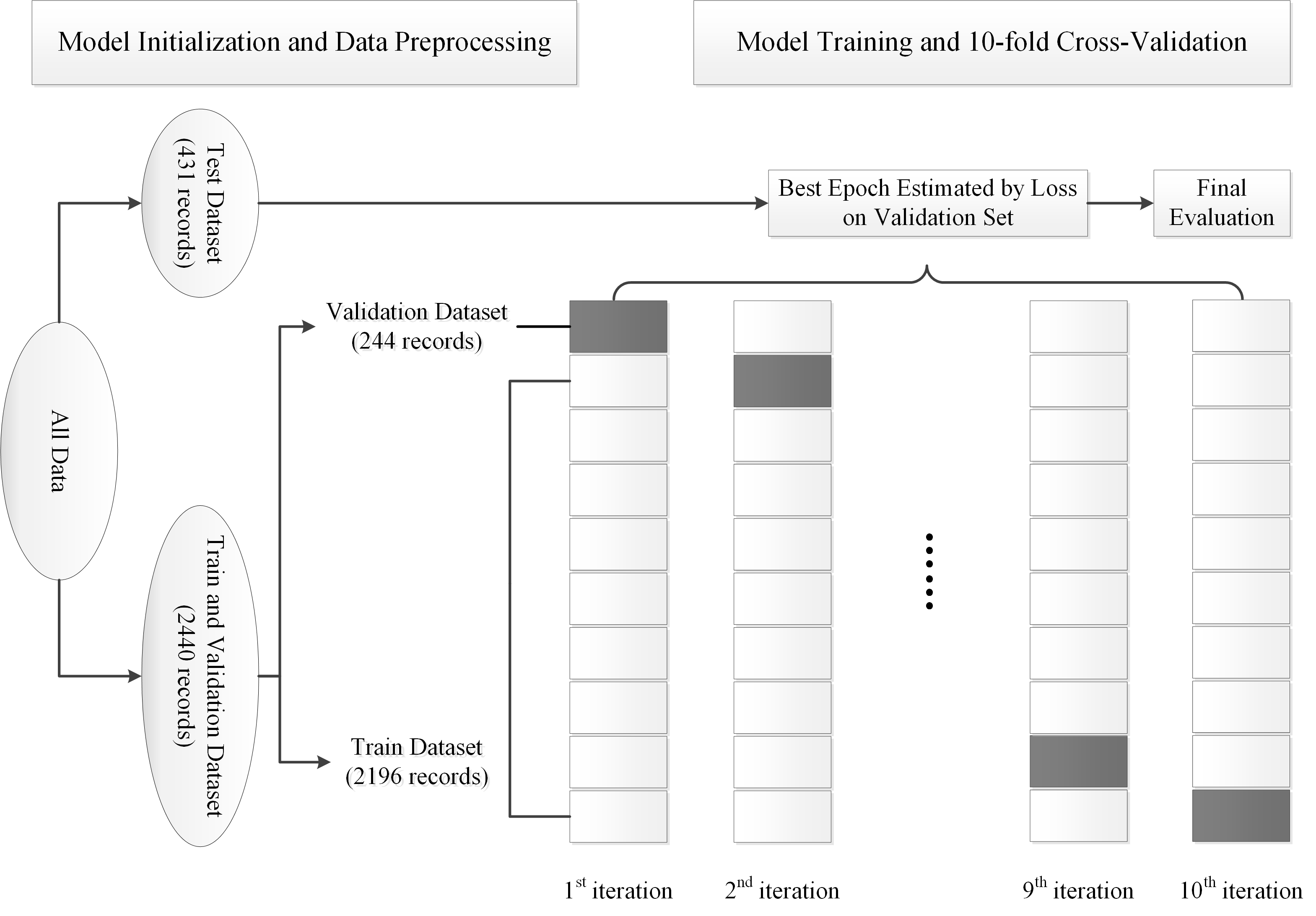}
			\caption{Experiment Process}
		\end{figure}
		
		In the training process of GNNWR, we use the RMSE as the loss function. We used the more popular Adam optimizer and achieved better results than the stochastic gradience descent used in GNNWR's previous practice. When the loss function of the validation set grows or remains constant beyond a certain number of iterations, the model is considered to be overfitted and the neural network computation is automatically stopped. For a given set of hyperparameters, 10 models can be generated on a randomly selected 10-fold data set (a total of 2440 items, accounting for 85\%) with 9 folds selected as the training set and the remaining 1 fold as the validation set. Secondly, by summing the loses of these ten models, the total model loses corresponding to the hyperparameters can be obtained. Finally, the hyperparameter of the model with the lowest mean value of loss was selected as the best, and the GNNWR it generated will be compared with the other two schemes.
		
		After several trials, in the hyperparameters, the value of learning rate is $10^{-2.95}\approx0.00112, \beta_{1}=0.8, \beta_{2}=0.999, \text{batch size}=128$. Percentage of loss at Dropout layer is 0.9; Epoch has a maximum number of iterations of 90,000. After comparing the results of several neural network hidden layers, the results are shown in Table \ref{layers}. Note that the data used to calculate the mean squared error here are derived from normalized house prices, so the order of magnitude is different from the analysis below.
		
		\begin{table}[H]
			\centering
			\small
			\begin{tabular}{|c|c|c|c|}
				\hline
				\begin{tabular}[c]{@{}c@{}}Structure of \\ Hidden Layers\end{tabular}        & \begin{tabular}[c]{@{}c@{}}Validation \\ Loss\end{tabular} & \begin{tabular}[c]{@{}c@{}}Train \\ Loss\end{tabular} & \begin{tabular}[c]{@{}c@{}}Test \\ Loss\end{tabular} \\ \hline
				\begin{tabular}[c]{@{}c@{}}{[}1024, 512, 256,\\  128, 64, 32{]}\end{tabular} & 0.006470                                                  & 0.002790                                             & 0.008867                                            \\ \hline
				{[}512, 128, 64, 16{]}                                                       & 0.006427                                                  & 0.0038040                                             & 0.008661                                            \\ \hline
				{[}512, 128, 32{]}                                                           & 0.006529                                                  & 0.0040193                                             & 0.008683                                            \\ \hline
				{[}256, 64, 16{]}                                                            & 0.006537                                                  & 0.0043795                                             & 0.008555                                            \\ \hline
				{[}256, 32, 8{]}                                                             & 0.006527                                                  & 0.0049904                                             & 0.008379                                            \\ \hline
				{[}256, 32{]}                                                                & 0.006567                                                  & 0.0046721                                             & 0.008992                                            \\ \hline
			\end{tabular}
		\caption{Loss for Different Structures}
		\label{layers}
		\end{table}

		After pre-experimental comparison, it was found that increasing the number of hidden layers not only greatly improved the fitting accuracy, but also did not significantly weaken the generalization effectiveness. Considering the number of neurons in the input and output layers, the article adopts a 6-layer neural network structure, containing 1 input layer, 4 hidden layers with 512, 128, 64, 16 neurons, and 1 output layer. The number of neurons in the input layer is the number of training samples, and the number of neurons in the output layer is the number of parameters of the linear regression model (the number of independent variables plus one).
		
\begin{figure*}[htbp]
			\centering
			\subfloat[The Decrease of AICc on Validation Set]{\includegraphics[width=0.3\textwidth]{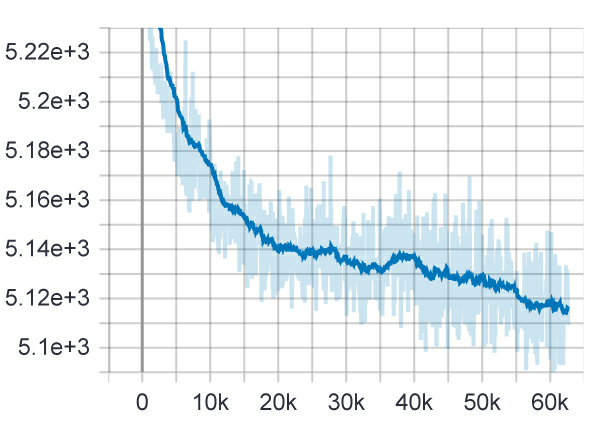}}
			\subfloat[The Decrease of AICc on Train Set]{\includegraphics[width=0.3\textwidth]{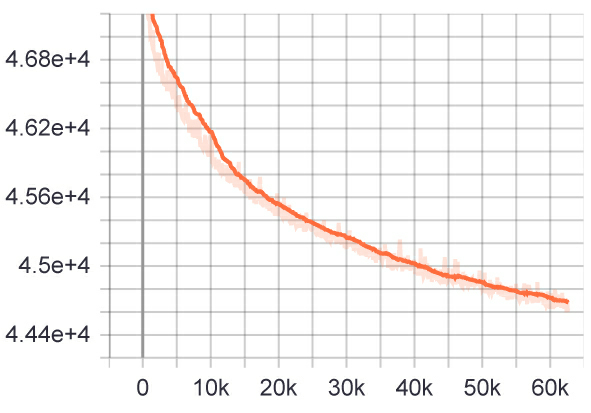}}
			\subfloat[The Decrease of Average Absolute Error]{\includegraphics[width=0.3\textwidth]{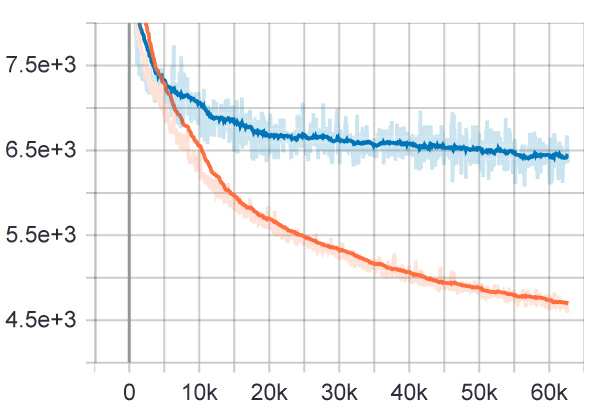}} \\
			\subfloat[The Decrease of Average Relative Error]{\includegraphics[width=0.3\textwidth]{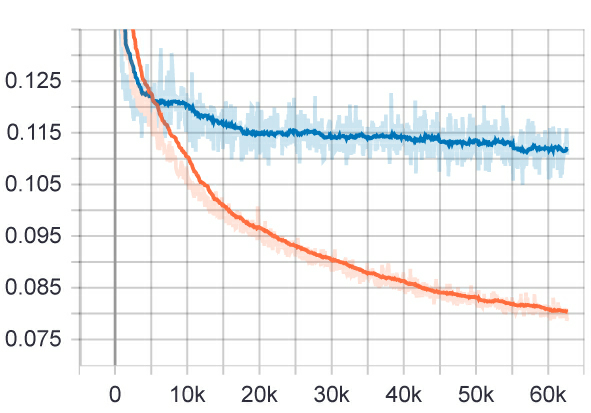}}
			\subfloat[The Increase of Determination Coefficient]{\includegraphics[width=0.3\textwidth]{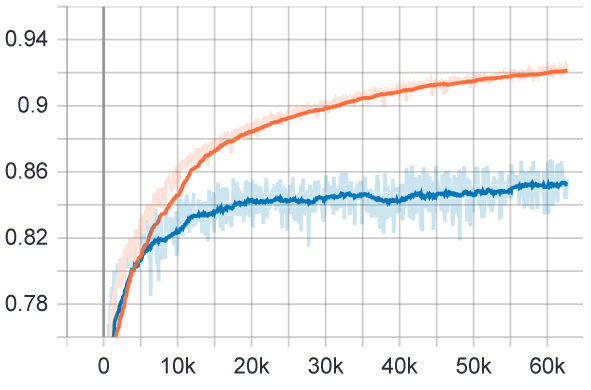}}
			\subfloat[The Decrease of Loss]{\includegraphics[width=0.3\textwidth]{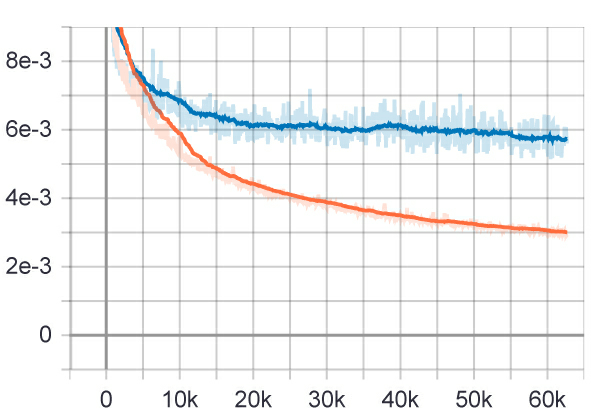}} \\
			
			\caption{Performance Variations for the Train (Orange) and Validation (Blue) Sets of GNNWR}
			\label{loss}
		\end{figure*}

		To further reflect the optimization in the iterations, Figure \ref{loss} shows the change in test indicators of one fold during model training. After running more than 30,000 epochs, if the loss on the validation set is not improved after 9000 epochs, the neural network training will be terminated.
		
		To fully demonstrate the superiority of GNNWR, three models, OLR, GWR, and GNNWR, are developed and compared on the Shenzhen house price dataset. Among them, GWR uses the golden search method to find the most suitable number of neighboring elements according to the AICc index.
		
		Since the NSS, QASP variables take values in small integers, the design matrix used in GWR modeling will show multicollinearity when the number of neighboring elements is small. Therefore, the search lower bound is set to 100, i.e., at least 100 neighboring elements are involved in the solution of the local regression coefficients.
		
		Through a simple pre-experiment, 2440 data are extracted for modeling and the remaining 431 data are used for testing, and it can be compared to find that bi-square significantly outperforms Gaussian among the two weighting kernel functions of GWR.The specific parameters are shown in the following Table \ref{GWR_kernel}. Therefore, the next comparisons in this paper all take bi-square as the weight kernel function of the GWR model.

		\begin{table*}[htbp]
			\small
			\centering
			\begin{tabular}{|c|cccccc|cc|}
				\hline
				\multirow{2}{*}{\begin{tabular}[c]{@{}c@{}}\\GWR Kernel \\ Type\end{tabular}} & \multicolumn{6}{c|}{Train}                                                                                                                                                                                                      & \multicolumn{2}{c|}{Test}                                                                        \\ \cline{2-9} 
				& \multicolumn{1}{c|}{R2}     & \multicolumn{1}{c|}{RMSE}     & \multicolumn{1}{c|}{MAE}      & \multicolumn{1}{c|}{MAPE/\%}  & \multicolumn{1}{c|}{AICc}    & \begin{tabular}[c]{@{}c@{}}Correlation \\ Coefficient\end{tabular} & \multicolumn{1}{c|}{R2}     & \begin{tabular}[c]{@{}c@{}}Correlation \\ Coefficient\end{tabular} \\ \hline
				Bi-square(105)                                                              & \multicolumn{1}{c|}{0.8861} & \multicolumn{1}{c|}{7655.203} & \multicolumn{1}{c|}{5623.454} & \multicolumn{1}{c|}{0.094994} & \multicolumn{1}{c|}{51842.0} & 0.941789                                                           & \multicolumn{1}{c|}{0.7935} & 0.892818                                                           \\ \hline
				Gaussian(101)                                                               & \multicolumn{1}{c|}{0.7471} & \multicolumn{1}{c|}{11408.08} & \multicolumn{1}{c|}{8372.857} & \multicolumn{1}{c|}{0.141284} & \multicolumn{1}{c|}{52790.3} & 0.865382                                                           & \multicolumn{1}{c|}{0.6120} & 0.783185                                                           \\ \hline
			\end{tabular}
		\caption{Performance for Different Kernel Types}
		\label{GWR_kernel}
		\end{table*}

		In the experiments, we used OLR as a comparison. Both OLR and GWR solutions are built on ArcGIS Pro. GNNWR built with TensorFlow 1.15.0 library under Python 3.6.13 kernel. The commonly used AICc criterion is chosen for GWR bandwidth optimization。
		
		We use these three modeling approaches with the help of 10-fold cross-validation to be able to build the model on the training set, use the results on the validation set to calculate each metric of the model, evaluate the generalization ability of the model, and exclude the influence of chance factors. Finally, the modeling result with the strongest generalization ability is selected for all three methods, and the predictive ability of the model is tested on the test set. For 2871 data, we extracted 431 of them (about 15\%) as the test set, and the remaining 2440 data were equally divided into 10 folds to participate in cross-validation, each containing 244 data (about 8.5\%).

		\section{Effectiveness Evaluation of GNNWR}
		
		\subsection{Dataset Analysis and Descriptive Statistics}
		The results of correlation analysis and descriptive statistics of Shenzhen house prices with the respective variable factors are shown in the Table \ref{Exploratory Analysis} below.
		
		\begin{table*}[htbp]
			\small
			\centering
			\addtolength{\leftskip} {-0.3cm}
			\begin{tabular}{|c|c|c|c|c|c|c|c|c|c|c|}
				\hline 
				Indicator & Price & AB & NPS & MF & GR & PR & SD & QAPS & NSS & DSS \\ \hline
				Mean & 62219.3 & 17.995 & 507.196 & 2.625 & 0.340 & 3.113 & 6586.1 & 3.704 & 1.769 & 930.5\\ \hline
				Maximum & 132000 & 51 & 5500 & 36.6 & 0.990 & 7.000 & 24967.2 & 4.5 & 8 & 25110.0\\ \hline
				Minimum & 16100 & 1 & 1 & 0 & 0.100 & 0.100 & 23.2 & 0 & 0 & 16.8\\ \hline
				Std. Dev. & 22986.5 & 7.138 & 647.509 & 1.888 & 0.130 & 1.438 & 4933.7 & 1.179 & 1.432 & 1838.5\\ \hline
				Correlation Coefficient & -	& -0.118 & 0.079 & 0.262 & 0.216 & 0.105 & -0.504 & 0.080 & 0.248 & -0.236\\ \hline
				Variation Coefficient & 2.707 & 2.521 & 0.783 & 1.391 & 2.620 & 2.164 & 0.749 & 3.143 & 1.235 & 0.506\\ \hline
				VIF & - & 1.622 & 1.243 & 1.227 & 1.122 & 1.150 & 1.167 & 1.204 & 1.365 & 1.136\\ \hline
				t-test p & - & 0 & 3.4E-08 & 0 & 0 & 0.0197 & 0 & 0.3208 & 0 & 0\\
				
				 \hline
				
			\end{tabular}
		\caption{Exploratory Analysis and Descriptive Statistics of the Experimental Dataset}
		\label{Exploratory Analysis}
		\end{table*}
		
		As can be seen from the table, ranked by the absolute values of the correlation coefficients in descending order, the variables are SD, MF, NSS, DSS, GR, AB, PR, QAPS, NPS. The variables positively correlated with house prices are MF, NSS, GR, PR, QAPS, NPS, and negatively correlated with SD, DSS, AB. The average price of residential housing in Shenzhen sampled in this dataset is $62219.33 yuan/m^2$. From the highest $132000 yuan/m^2$ to the lowest $16100yuan/m^2$, the value domain basically covers the reference price of second-hand housing transactions for all residential housing in Shenzhen at present. VIF measures the extent to which this independent variable is influenced by other independent variables, and since all of them are extremely close to 1, it can be found that the degree of multicollinearity of the data source is minimal.
		
		Ranked by coefficient of variation in descending order, the variables are QAPS, Price, GR, AB, PR, MF, NSS, NPS, SD, DSS. After normalization, QAPS is the variable with the greatest difference, while DSS varies the least among properties. In general, the price fluctuation is also relatively large, and the price difference between high-quality and low-quality properties is obvious, which more truly reflects the scarcity and non-renewable nature of land resources.
		
		It should be noted that this study also conducted hypothesis testing for each variable in the global regression equation using R language. For each variable, it is assumed that its coefficient is zero in the global regression equation, and a test statistic that satisfies the t-distribution when the hypothesis holds can be constructed. Correspondingly, the p-value can be calculated. The p-values of AB, MF, GR, SD, NSS, and DSS will be recorded as 0 because they are less than the accuracy threshold of $2.2\times10^{-16}$. The p-values of NPS were also very small and highly significant. It can also be found that the p-value of PR is not significant at the significance level of 0.01 and the p-value of QAPS is not significant at the significance level of 0.05 or even 0.1. If a global regression is used, these two independent variables should be excluded. However, two spatial statistical modeling methods, GNNWR and GWR, are used in this study, and the significance of each variable in this model can be re-tested with the help of the F2 statistic in this paper. According to the analysis of non-stationarity diagnostics in 3.3, both variables are highly significant, when the coefficients of the linear model are allowed to vary with geographic coordinates. This proves the superiority of the spatial statistical modeling approach from another side.
		
		\subsection{Comparison of Indicators of House Price Valuation Models}
		
		The evaluation of the house price valuation model examines both the ability to fit on the training set and to predict on the test set. We stochastically divides 2871 house price records into the train set and validation set with 2440 records as 10 folds, and 431 records remnant working as the test set.
		
		We evaluated all parameters of the model using parameters such as coefficient of R$^2$, RMSE, MAE, MAPE, AICc and Pearson correlation coefficient. For the dataset generated after the 10-fold crossover, the validation sets are merged and the following results are obtained.
		
		\begin{table*}[htbp]
			\small
			\centering
			\begin{tabular}{|c|c|c|c|c|c|c|}
				\hline
				Model & R2       & RMSE     & MAE      & MAPE     & Mean Err. & Pearson Cor. Coe. \\ \hline
				GNNWR & 0.840177 & 9069.561 & 6558.630 & 0.111965 & 27.88808  & 0.916637          \\ \hline
				GWR   & 0.788728 & 10427.68 & 7581.746 & 0.128538 & -73.9177  & 0.888123          \\ \hline
				OLS   & 0.432101 & 17096.31 & 13003.76 & 0.228767 & -5.60228  & 0.657404          \\ \hline
			\end{tabular}
		\caption{Indicators of GNNWR, GWR and OLS on Merged Validation Set}
		\end{table*}
		
		Clearly, these data confirm the greatness of the GNNWR model. The worst prediction comes from the OLS model, which has the lowest R$^2$ and the highest prediction error measured by RMSE, MAE and MAPE. Because of the severe spatial non-stationarity, the OLS model is difficult to detect the intrinsic relations and spatial fluctuations between house price and other independent variables. Compared with GWR model, the RMSE of GNNWR model declines about 13.0\%, and the MAE of GNNWR model declines about 13.5\%. Other indicators, like R2 and MAPE, also make the superiority of GNNWR model clear. Additionally, the mean residual error of GNNWR model is much lower than GWR model’s with a 62.2\% reduction, which means that the prediction of GNNWR model has a greater unbiasedness than GWR's on this dataset. In short, we can deduce that the GNNWR model gains a notable progress on the generalization ability.
		
		To be more specific, we can compare the indicators of the GWR and the GNNWR models in each process of modeling on 10 train sets. There parameters reflect the fitting quality of modeling process. In Table \ref{indicators}, the Train set of 0 means that data set 0 is excluded and the 1, 2, ..., 9 data sets are selected, and so on.
		
		\begin{table*}[htbp]
			\small
			\centering
			\begin{tabular}{|c|l|c|c|c|c|c|c|}
				\hline
				Train Set          & \multicolumn{1}{c|}{Model} & R2     & RMSE    & MAE     & MAPE     & Pearson Cor. Coe. & AICc     \\ \hline
				\multirow{2}{*}{0} & GNNWR                      & 0.9130 & 6665.74 & 4907.16 & 0.084455 & 0.955890          & 44935.93 \\ \cline{2-8} 
				& GWR(108)                   & 0.8806 & 7810.28 & 5722.90 & 0.096476 & 0.938932          & 46711.89 \\ \hline
				\multirow{2}{*}{1} & GNNWR                      & 0.9145 & 6698.92 & 4945.16 & 0.084418 & 0.956290          & 44923.25 \\ \cline{2-8} 
				& GWR(101)                   & 0.8881 & 7662.40 & 5624.61 & 0.095042 & 0.942811          & 46714.44 \\ \hline
				\multirow{2}{*}{2} & GNNWR                      & 0.9168 & 6516.65 & 4849.38 & 0.084081 & 0.957767          & 44799.69 \\ \cline{2-8} 
				& GWR(101)                   & 0.8835 & 7713.28 & 5663.39 & 0.095898 & 0.940476          & 46751.67 \\ \hline
				\multirow{2}{*}{3} & GNNWR                      & 0.9068 & 6923.85 & 5118.84 & 0.087057 & 0.952315          & 45066.73 \\ \cline{2-8} 
				& GWR(101)                   & 0.8887 & 7566.31 & 5594.39 & 0.094180 & 0.943155          & 46666.59 \\ \hline
				\multirow{2}{*}{4} & GNNWR                      & 0.9180 & 6489.22 & 4784.29 & 0.080781 & 0.958206          & 44776.13 \\ \cline{2-8} 
				& GWR(101)                   & 0.8842 & 7711.99 & 5656.98 & 0.095410 & 0.940830          & 46748.33 \\ \hline
				\multirow{2}{*}{5} & GNNWR                      & 0.9109 & 6777.58 & 4990.82 & 0.086299 & 0.954517          & 44972.38 \\ \cline{2-8} 
				& GWR(101)                   & 0.8849 & 7702.18 & 5664.53 & 0.095798 & 0.941220          & 46744.85 \\ \hline
				\multirow{2}{*}{6} & GNNWR                      & 0.9175 & 6538.53 & 4796.90 & 0.082787 & 0.958058          & 44850.11 \\ \cline{2-8} 
				& GWR(106)                   & 0.8814 & 7839.84 & 5715.71 & 0.096394 & 0.939293          & 46751.99 \\ \hline
				\multirow{2}{*}{7} & GNNWR                      & 0.9183 & 6529.28 & 4827.74 & 0.081521 & 0.958488          & 44795.06 \\ \cline{2-8} 
				& GWR(101)                   & 0.8871 & 7675.50 & 5658.68 & 0.095299 & 0.942348          & 46730.64 \\ \hline
				\multirow{2}{*}{8} & GNNWR                      & 0.9077 & 6868.49 & 5104.66 & 0.087495 & 0.953352          & 45038.13 \\ \cline{2-8} 
				& GWR(104)                   & 0.8825 & 7749.99 & 5713.72 & 0.096287 & 0.939906          & 46735.51 \\ \hline
				\multirow{2}{*}{9} & GNNWR                      & 0.9082 & 6814.82 & 4998.50 & 0.086338 & 0.953331          & 45025.80 \\ \cline{2-8} 
				& GWR(107)                   & 0.8796 & 7802.49 & 5714.57 & 0.096653 & 0.938439          & 46719.31 \\ \hline
			\end{tabular}
		\caption{Indicators of GWR and GNNWR on Train Sets}
		\label{indicators}
		\end{table*}
		
		It should be noted that the number after GWR refers to the number of most suitable neighboring elements selected based on the AICc value. Since the training set is slightly different, the most appropriate number of neighboring elements is re-picked each time the GWR model is built.
		
		For all of these 10 data sets, GNNWR models have completely beaten GWR models no matter we utilize AICc, RMSE, R2 or Pearson correlation coefficient as a judge. The evident advantage on AICc reveals that the GNNWR model not only provides a better prediction about house price, but also applies a more accurate space weight matrix without much more complexity. In contrast, the GWR model has to face the overfitting problem, which makes the correctness of the predictions on the validation sets slump. To sum up, the GNNWR model producing a more capable kernel function than any GWR models, performs most outstandingly in catching spatial heterogeneity details, estimating spatial weight and predicting dependent variables.
		
		Furthermore, we can judge the generalization ability by predicting the test set. In this study, we use the models with the best generalization ability to compare. Both of the GNNWR and the GWR models perform best when we opt the validation set as dataset 4, and the other indicators are shown in the next Table \ref{indicators test}.
		
		\begin{table*}[htbp]
			\small
			\centering
			\begin{tabular}{|c|c|c|c|c|c|c|}
				\hline
				Test Set & R2       & RMSE     & MAE      & MAPE     & Mean Err. & Pearson Cor. Coe. \\ \hline
				GWR      & 0.790389 & 11195.01 & 7912.005 & 0.122266 & 911.3839  & 0.891319          \\ \hline
				GNNWR    & 0.817178 & 10455.19 & 7108.715 & 0.109174 & 1393.691  & 0.905834          \\ \hline
			\end{tabular}
		\caption{Indicators of GWR and GNNWR on Test Sets}
		\label{indicators test}
		\end{table*}
		
		Compared with the GWR model, the GNNWR model has an explicit superiority about predicting the test dataset. The MAE slumps 10.2\% and the MAPE descents 10.7\%, which are practical for real estate agency to have a better estimation. The RMSE reduces 6.6\%, the R2 and the Pearson correlation coefficient has improved as well and the mean error has increased. 
		
		\subsection{Spatial Non-stationarity Diagnosis of House Price Regression Relationship}
		
		Based on the spatial heterogeneity diagnostic indicators discussed above, the results based on GNNWR can be analyzed in two parts.
		
		First, it is possible to identify whether the model results have a relatively significant spatial non-smoothness. In the ten-fold data, the prediction effect parameters of each GNNWR model in the validation set are as Table \ref{prediction}.
		
		\begin{table*}[htbp]
			\small
			\centering
			\begin{tabular}{|c|c|c|c|c|c|}
				\hline
				Validation Set & R2       & RMSE     & MAE      & MAPE     & Pearson Cor. Coe. \\ \hline
				0              & 0.836963 & 9454.353 & 6876.564 & 0.115830 & 0.915256          \\ \hline
				1              & 0.821430 & 8697.789 & 6277.030 & 0.105508 & 0.907206          \\ \hline
				2              & 0.855282 & 8930.441 & 6546.944 & 0.109760 & 0.924835          \\ \hline
				3              & 0.812268 & 9812.707 & 6643.969 & 0.112837 & 0.901745          \\ \hline
				4              & 0.871563 & 8189.907 & 5944.871 & 0.104658 & 0.933591          \\ \hline
				5              & 0.837077 & 9098.732 & 6577.607 & 0.111918 & 0.915711          \\ \hline
				6              & 0.840490 & 8783.315 & 6620.163 & 0.113857 & 0.917376          \\ \hline
				7              & 0.813833 & 9034.993 & 6549.102 & 0.115053 & 0.902763          \\ \hline
				8              & 0.840247 & 9332.620 & 6836.936 & 0.117101 & 0.916725          \\ \hline
				9              & 0.854961 & 9260.348 & 6713.113 & 0.113130 & 0.925135          \\ \hline
			\end{tabular}
		\caption{Prediction Performace of 10 GNNWR Models on Each Validation Set}
		\label{prediction}
		\end{table*}
		
		Using RMSE as the index, the best fitting model (model 4) and the worst fitting model (model 3) were selected for hypothesis testing. The hypothesis testing parameters were calculated from the previous derivation as the following Table \ref{F1}.
		
		\begin{table*}[htbp]
			\small
			\centering
			\begin{tabular}{|c|c|c|c|c|c|}
				\hline
				F1 Hypothesis Test & F1  & $\sigma_1$       & $\sigma_2$        & Distribution    & Significant Level \\ \hline
				Best Fitting Model    & 0.071602 & 4439.122 & 4871141  & F(4.0454, 2186) & 1E-2              \\ \hline
				Worst Fitting Model   & 0.117877 & 3118.766 & 804246.6 & F(12.094, 2186) & 1E-4              \\ \hline
			\end{tabular}
		\caption{F1 Hypothesis Test}
		\label{F1}
		\end{table*}
		
		Following $F_{1}$ value, we can deduce the p value so that the hypothesis establishes by calculating the F distribution. It is notably that the hypothesis is rejected and it is significant that there is severe spatial non-stationarity when modeling Shenzhen house price.
		
		Next, we could analyze the significance for each independent variable. To every independent variable, the null hypothesis is that the coefficient of this variable is a constant. It is to be noted that this hypothesis includes another hypothesis which assumes the coefficient of this variable is 0 everywhere. Therefore, the p value of $F_{2}$ can reject both of the hypotheses if it is tiny enough. All of the details are shown in the Table \ref{F2}.
		
		\begin{table*}[htbp]
			\small
			\centering
			\addtolength{\leftskip} {-0.3cm}
			\begin{tabular}{|c|c|c|c|c|c|c|c|c|c|c|c|}
				\hline
				Model                                                                              & Variable                                                     & Intercept & AB     & NPS    & MF     & SD     & GR     & PR     & QAPS   & NSS    & DSS    \\ \hline
				\multirow{4}{*}{\begin{tabular}[c]{@{}c@{}}Best \\ Fitting \\ Model\end{tabular}}  & F Value                                                      & 614.58    & 234.48 & 381.14 & 562.52 & 537.77 & 503.31 & 385.10 & 418.17 & 646.79 & 502.37 \\ \cline{2-12} 
				& $\gamma_1$                                                           & 0.0198    & 0.0893 & 0.4068 & 0.4108 & 0.4200 & 0.4560 & 0.5951 & 0.6102 & 0.6753 & 0.6692 \\ \cline{2-12} 
				& $\gamma_2$                                                           & 0.0004    & 0.0057 & 0.1095 & 0.1085 & 0.1108 & 0.1120 & 0.1629 & 0.1595 & 0.1789 & 0.1824 \\ \cline{2-12} 
				& \begin{tabular}[c]{@{}c@{}}Significant \\ Level\end{tabular} & 1E-10     & 1E-08  & 1E-09  & 1E-10  & 1E-10  & 1E-10  & 1E-10  & 1E-11  & 1E-12  & 1E-11  \\ \hline
				\multirow{4}{*}{\begin{tabular}[c]{@{}c@{}}Worst \\ Fitting \\ Model\end{tabular}} & F2                                                     & 1017.10   & 471.84 & 573.85 & 872.23 & 845.57 & 774.72 & 344.85 & 367.06 & 560.84 & 432.05 \\ \cline{2-12} 
				& $\gamma_1$                                                           & 0.0289    & 0.1302 & 0.5264 & 0.5279 & 0.5338 & 0.5990 & 1.3826 & 1.3973 & 1.5010 & 1.4639 \\ \cline{2-12} 
				& $\gamma_2$                                                           & 0.0008    & 0.0115 & 0.1760 & 0.1741 & 0.1758 & 0.1807 & 0.9202 & 0.9138 & 0.9341 & 0.8760 \\ \cline{2-12} 
				& \begin{tabular}[c]{@{}c@{}}Significant \\ Level\end{tabular} & 0         & 1E-04  & 1E-04  & 1E-05  & 1E-05  & 1E-05  & 1E-04  & 1E-04  & 1E-04  & 1E-04  \\ \hline
			\end{tabular}
		\caption{F2 Hypothesis Test}
		\label{F2}
		\end{table*}
		
		Evidently, every independent variable has significant influence on house price, and each of their influence varies a lot among the region. Hence it has also been proved that all of them have significant spatial non-stationarity. What’s more, this simple comparison also hints that a better model may require a higher spatial non-stationarity estimation on variables and a lower spatial non-stationarity estimation on the intercept.
		
		\section{Comparison and Analysis}
		
		\subsection{Comparison of Prediction Performace of House Price Valuation Models}
\begin{figure}[H]
			\centering
			\subfloat[GWR-OLS]{\includegraphics[width=0.4\textwidth]{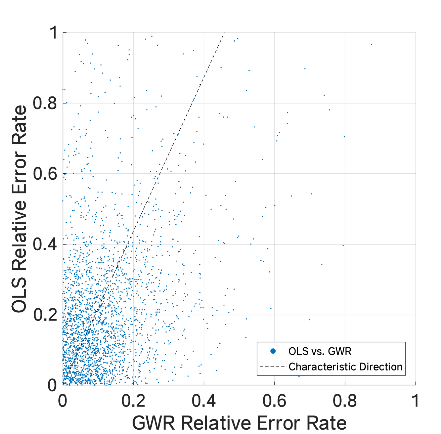}}\\ 
			\subfloat[GNNWR-OLS]{\includegraphics[width=0.4\textwidth]{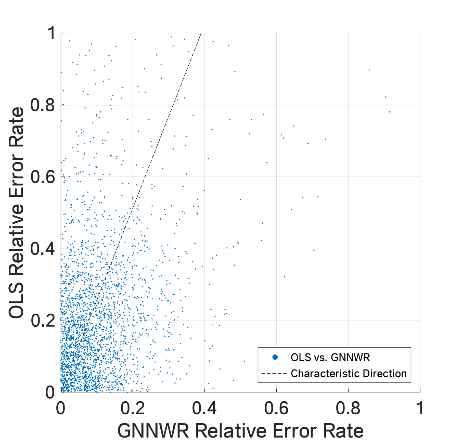}}\\ 
			\subfloat[GNNWR-GWR]{\includegraphics[width=0.4\textwidth]{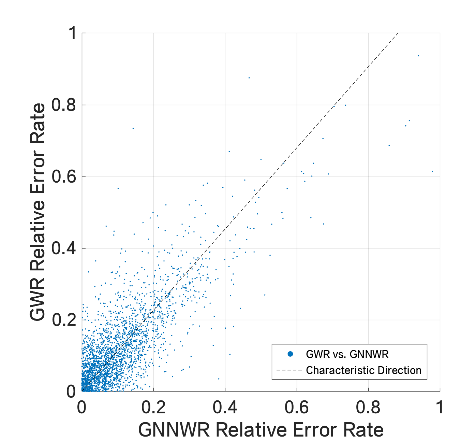}}\\ 
			\caption{Point Clouds and Feature Directions of Error Rates}
			\label{plots}
		\end{figure}

		The relative error rate of each prediction is calculated on the validation set and test set, which can be plotted as the following scatter plot in Figure \ref{plots}. Among them, the feature directions of the point cloud can be found according to the method of principal component analysis (PCA), and are plotted on the graph using black dashed lines. It should be noted that we use the same range of axes when plotting the point cloud in order to make the comparison clearer, and there are no more than 5\% of points outside this range that are not shown.

		It is not difficult to find that GWR and GNNWR have great superiority over the OLS models. The point set is densely distributed close to the y-axis, indicating that most of the locations that cannot be well predicted by OLS can be more effectively and accurately predicted by spatial statistical models. The idea of local linear regression can effectively reduce the prediction error. Comparing GWR with GNNWR, we can find that the feature direction lies above y=x, i.e. GNNWR can reduce the prediction error of GWR model at the same location by a certain proportion.
		
\begin{figure}[H]
			\centering
			\subfloat[On Merged Validation Set]{\includegraphics[width=0.45\textwidth]{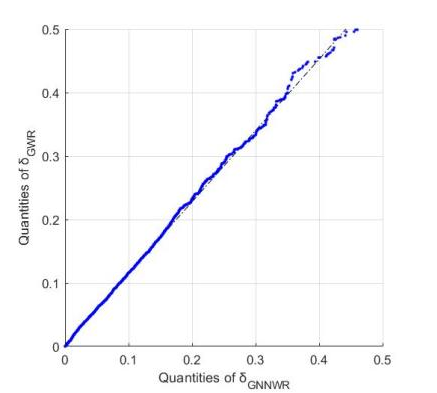}} \\
			\subfloat[On Test Set]{\includegraphics[width=0.45\textwidth]{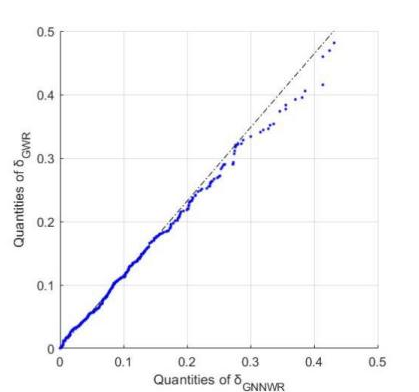}} 
			\caption{Q-Q Plots of Relative Error Rates}
			\label{Q-Q}
		\end{figure}

		Further, we compare the two models by a Q-Q plot and a histogram chart as Figure \ref{Q-Q}, which are plotted by Matlab.
		
		Again, no more than 5\% of the points are not shown outside this range. Ordering the relative error rates, it can be found that the relationship between the k$^{th}$ value on the validation set is approximately $\delta^{(k)}_{GWR}=1.123 \delta^{(k)}_{GNNWR}+0.0033$. The relationship between the k$^{th}$ value on the test set is approximately $\delta^{(k)}_{GWR}=1.160 \delta^{(k)}_{GNNWR}+0.0003$. These reference lines that represent the theoretical distribution have a clear deviation with $y=x$, which enable us to confirm the superiority of GNNWR models.
		
\begin{figure}[H]
			\centering
			\subfloat[On Merged Validation Set]{\includegraphics[width=0.5\textwidth]{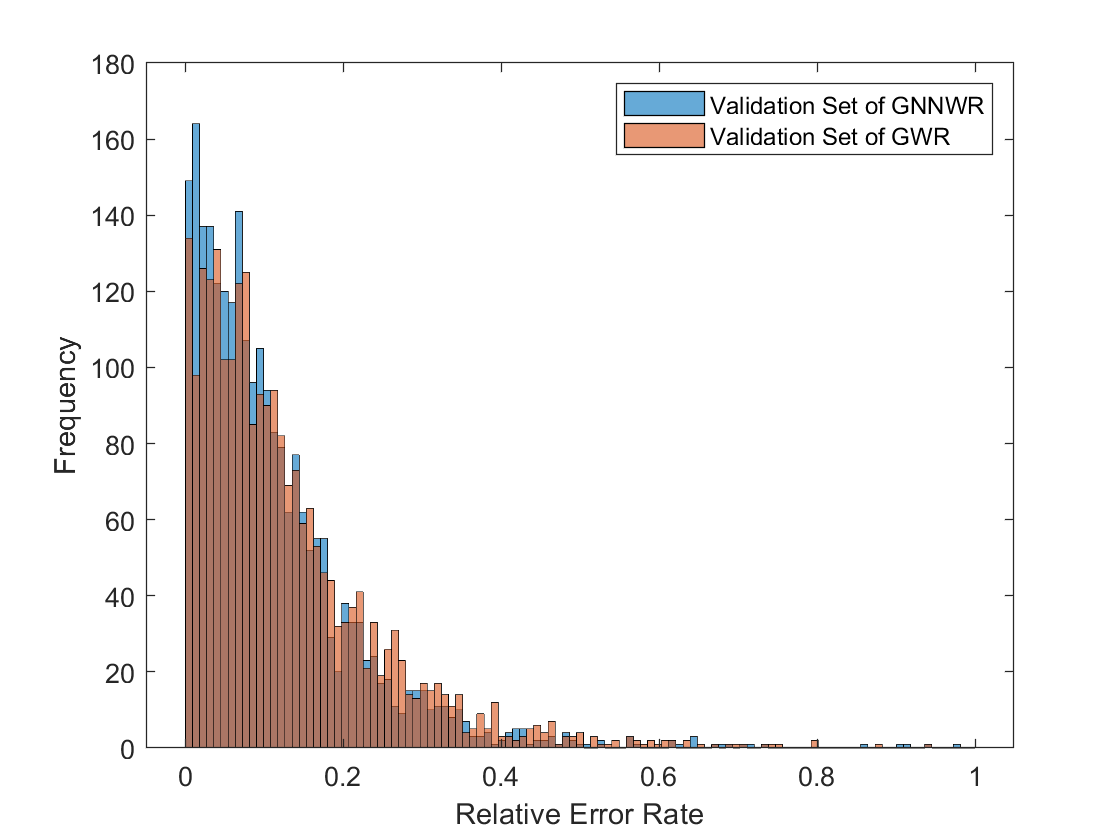}} \\
			\vspace{-0.1in}
			\subfloat[On Test Set]{\includegraphics[width=0.5\textwidth]{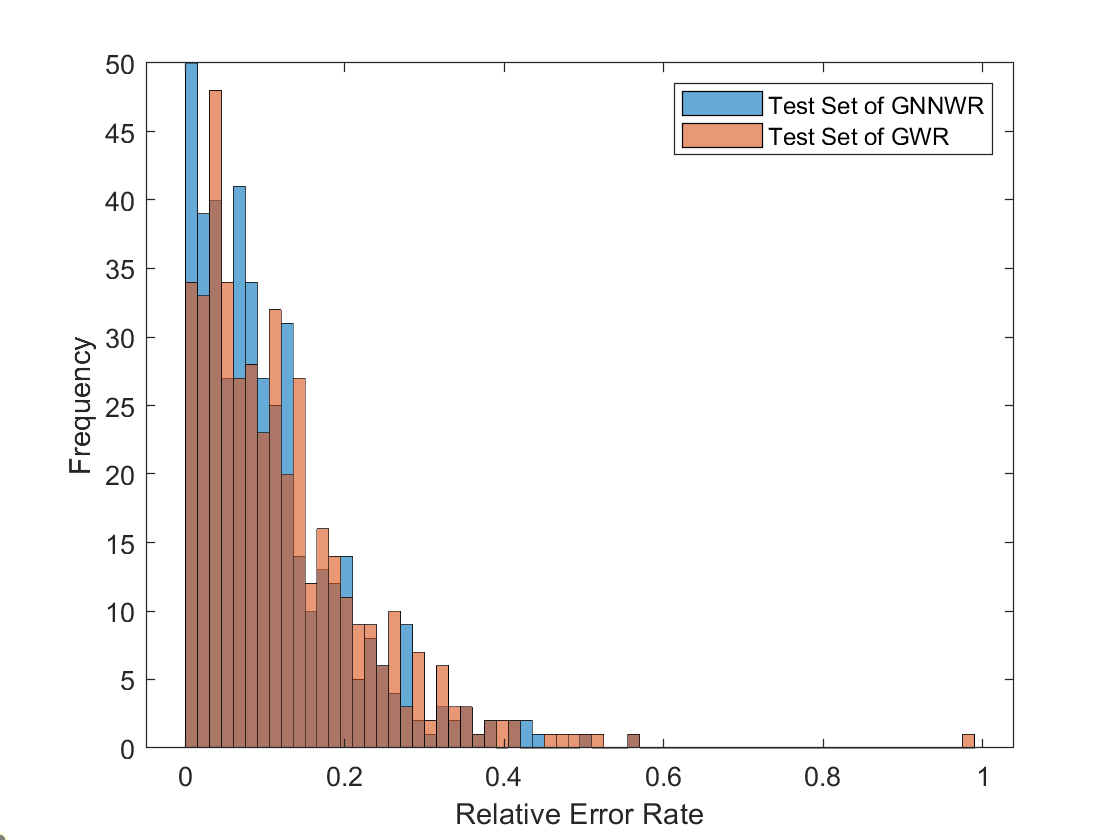}} 
			\caption{Histograms of Relative Error Rates}
			\label{histogram}
		\end{figure}

		A comparison of the histograms as Figure \ref{histogram} still gives clear results. On the validation set, taking the histogram horizontal coordinates between [0,1] and bin width of 0.09, it can be found that 9 of the 11 bins with error rate less than or equal to 9.9\% have more data from GNNWR model.This trend is also evident in the test set. Setting the histogram horizontal coordinates between [0, 1] and bin width taking 0.15, similarly it can be found that five of the seven bins with error rate less than or equal to 10.5\% have more data from the GNNWR model.
		
\begin{figure}[H]
			\centering
			\includegraphics[width=0.5\textwidth]{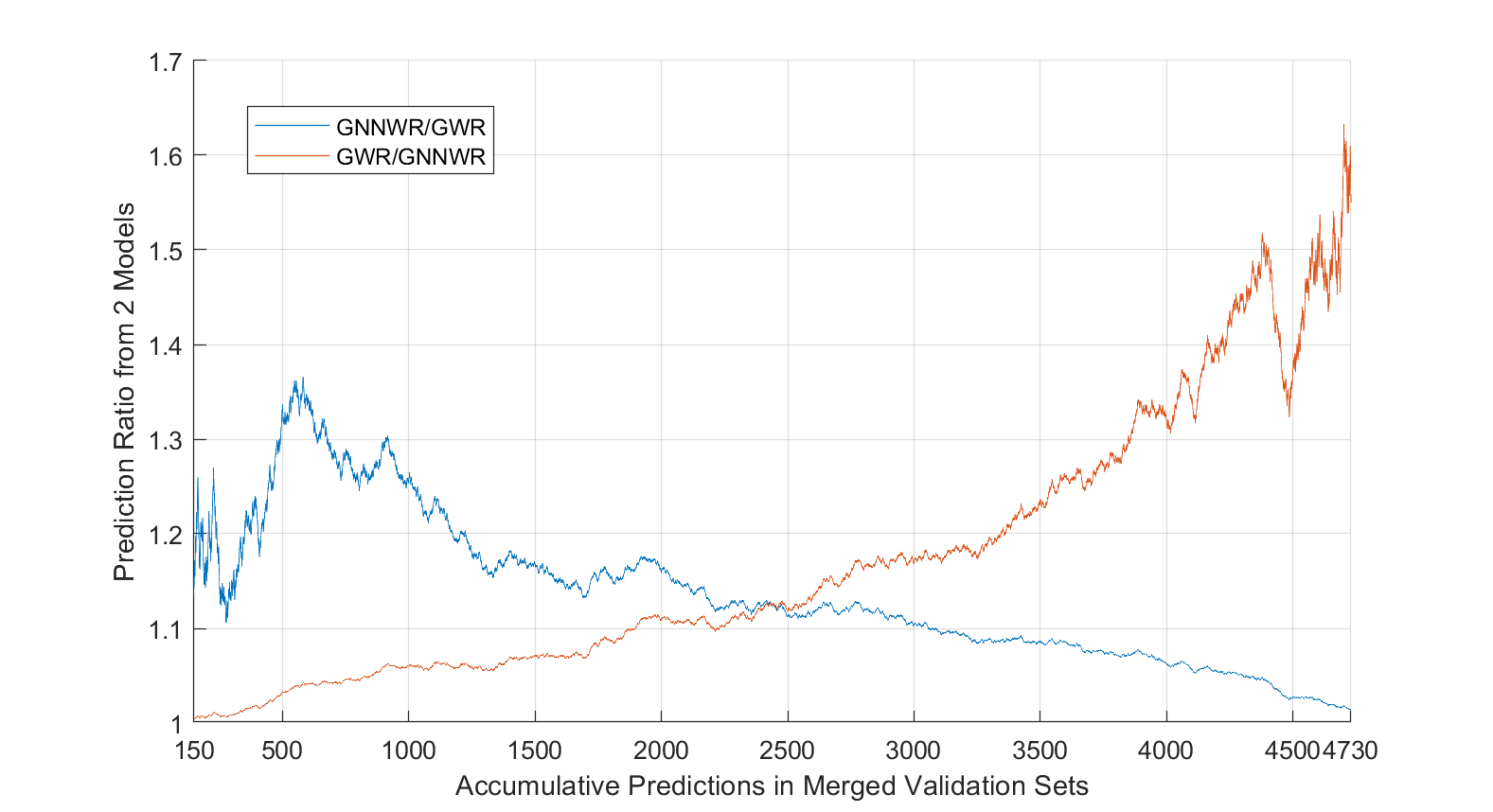}
			\caption{The Ratio of the Top N Best/Worst Predictions from 2 Models}
			\label{ratio}
		\end{figure}

		Besides, we can combine the prediction data of validation set from both models as Figure \ref{ratio}. The numbers of data in both sets which are below certain value can be calculated, and the ratio of two numbers can be plotted as blue line on the graph. The ratio of the number of predicted data from GWR to the number of data from GNNWR when the statistical error rate is above a certain value can be plotted as the orange line on the graph. For all data (two times of predictions on 2440 records) with a relative error rate of less than 0.203, the predictions from GNNWR are 1.34 times higher than those from GWR. In contrast, among all data with error rates higher than 0.37, the predictions from GWR are as much as 1.62 times higher than those from GNNWR. In conclusion, the predictions from GNNWR account for more of the high-precision predictions and the predictions from GWR account for more of the high-error predictions.
		
		Comparing with other literature, it can be found that another study also supports the conclusion that GWR can significantly reduce the prediction error compared to OLS models, indicating that spatial heterogeneity exists. In another study on Shenzhen house prices, the authors used the GWR model to increase the R$^{2}$ from 0.56 to 0.79.\cite{ref11} Some simple AI models, such as decision tree models, can even predict worse than OLS if they are not designed properly.\cite{ref35} In a separate study comparing the OLS model with multiple models, the best Stepwise and tuned SVM model reduced the RMSE by 25\%, the polynomial regression model reduced the RMSE by 8.3\%, and even the optimal simple neural network selected from the 1-3 hidden layers increased the RMSE by 66\%.\cite{ref36} Since the 1990s, scholars have been trying to use ordinary neural network models to predict house prices and compare them with OLS models. Some studies have demonstrated the superiority of the neural network approach, but others have found that there is no great need to use neural networks. Considering the 47\% reduction in RMSE metrics compared to OLS in this study, it is easy to see that simply using complex functions trained by neural networks to approximate the training data set does not improve the prediction accuracy, and that a GWR-based framework can best capture information on the geographic distribution. These indicate that accurate estimation of spatial heterogeneity is extremely necessary.
		
		\subsection{Analysis of Each Variable on House Prices}
		
		The GNNWR model is based on the structure of linear regression, where different weights are assigned to different variables based on the location of the property to capture spatial heterogeneity. For the ten-fold dataset obtained in this study, the weights of different independent variables at each prediction point can be visualized and output after merging the validation sets among them. This section focuses on the analysis of the significance of these weights.
		
		\begin{table*}[htbp]
			\small
			\centering
			\begin{tabular}{|c|c|c|c|c|c|c|c|c|c|c|}
				\hline
				Weight of Variables & AB     & NPS    & MF     & GR     & PR     & SD     & QASP   & NSS    & DSS    & Intercept \\ \hline
				Mean                & -0.280 & 0.170  & 0.458  & 0.114  & 0.002  & -0.474 & 0.021  & 0.033  & -0.160 & 0.508     \\ \hline
				Maximum             & 0.612  & 1.101  & 2.675  & 0.836  & 0.320  & 0.383  & 0.179  & 0.701  & 6.763  & 1.486     \\ \hline
				Minimum             & -1.450 & -0.420 & -1.322 & -0.230 & -0.191 & -2.035 & -0.055 & -0.868 & -4.627 & -0.018    \\ \hline
				Std. Dev.           & 0.195  & 0.179  & 0.609  & 0.108  & 0.057  & 0.253  & 0.026  & 0.187  & 0.820  & 0.208     \\ \hline
			\end{tabular}
		\caption{Descriptive Statistics of Weights of Variables}
		\label{weight}
		\end{table*}
		
		Since the data are pre-normalized when they enter GNNWR training, the values here can be compared directly in Table \ref{weight}. As can be seen from the mean values, the degree to which each independent variable affects house prices is different, and after taking the absolute values, they are SD, MF, AB, NPS, DSS, GR, NSS, QASP and PR, from the highest to the lowest. After accounting for spatial heterogeneity, the effect of NPS and AB on house prices is larger than that estimated using the correlation coefficient, and the effect of NSS on house prices is smaller than that estimated using the correlation coefficient. However, the positive and negative correlations of house prices are not violated, still MF, NPS, GR, NSS, QASP, PR are positively correlated with house prices and SD, AB, DSS are negatively correlated with house prices. The standard deviations of these weights were compared, from highest to lowest, as DSS, MF, SD, AB, NSS, NPS, GR, PR, and QASP. That is, public transportation conditions represented by DSS have greater spatial heterogeneity and school district conditions represented by QASP have less spatial heterogeneity, which is in line with the majority’s intuition. It can be speculated that since the value of quality educational resources is similar for residents in all parts of the city, the school district factor contributes to house prices with a more stable weight in all parts of Shenzhen. It can also be presumed that the distance to the subway station is not so important for residences living in the CBD or closely nearby the subway entrances. However, in the ordinary residential areas of the city and suburbs lacking wealthy people, the distance to the subway station is quite important. In-depth analysis requires specific distributions about the weights of each variable, as shown by Figure a-z. These figures are based on the natural breakpoint method with inconsistent color ranges for different subplots, and the boundaries around 0 are fine-tuned to show positive and negative correlation features. Due to the small standard deviation, the data of PR and QASP were classified into 6 levels only, while all other variables were classified into 8 levels. Overall, the modeling results based on ten different training sets are smooth, with few mutations and outliers in geographic proximity. They are quite consistent when making predictions for the weight distribution.
		
		\subsubsection{Analysis of Intercept Term Distribution}
		
		\begin{figure}[H]
			\centering
			\includegraphics[width=0.5\textwidth]{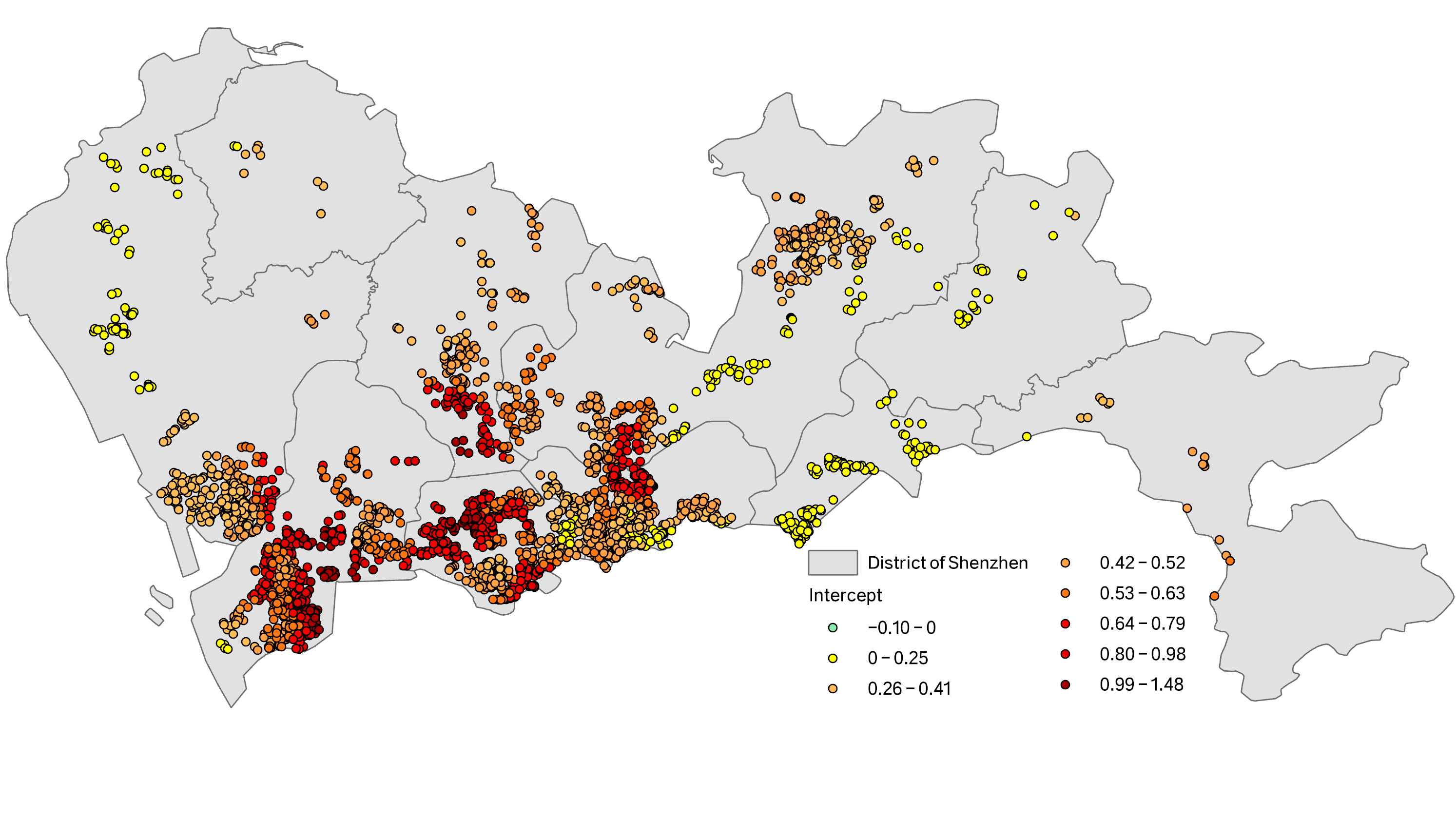}
			\caption{Intercept Distribution}
		\end{figure}
		
		The meaning of the intercept term is the inherent premium of the house after considering all the effects from the independent variables. It can be found by the graph that the reference prices for second-hand housing transactions introduced by the government gives the highest inherent premium to the coast of Nanshan District with the Houhai as the core, and the middle of Futian District with the east shore of Xiangmi Lake as the core. Because of the scarcity of premium locations, the market must be more frantic to capture this information and give higher premiums. In 2020, the highest residential transaction price of these two sites at \$50,000-\$70,000 per square meter continues to set a new record for housing prices in Shenzhen, while the relatively more marginal residences return to \$10,000-\$20,000. At this level, the reference price from government succeeds in perceiving the inherent premium distribution and narrowing the gap between inherent premiums. The GNNWR model is similarly able to provide an accurate estimate of the premium inherent in each block based on the reference price.
		
		\subsubsection{Analysis of Endogenous Variable}
		
		The property endogenous variables used in this model are MF, AB, NPS, GR and PR, in descending order of influence.
		
		\begin{figure}[H]
			\centering
			\includegraphics[width=0.5\textwidth]{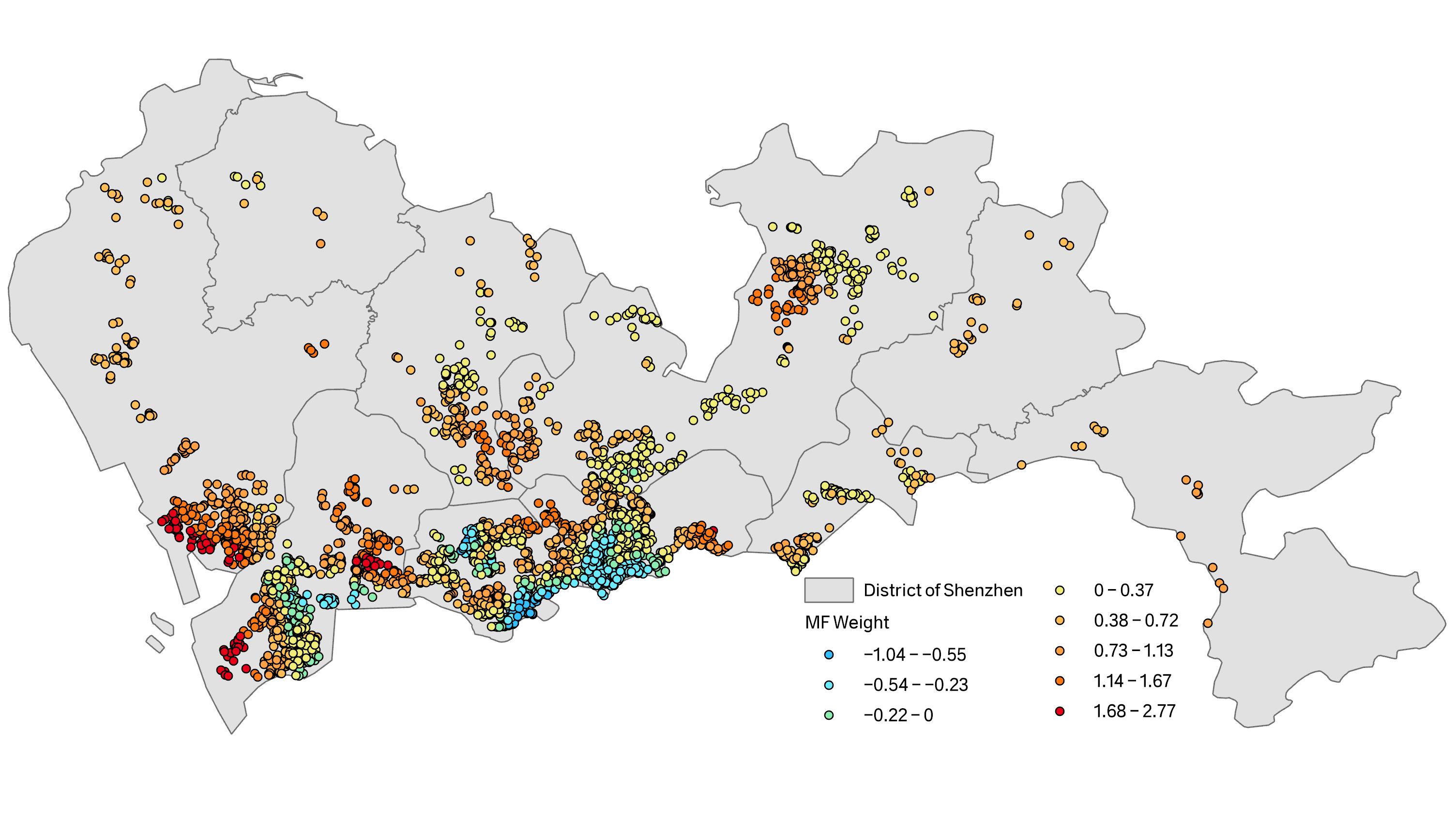}
			\caption{MF Weight Distribution}
		\end{figure}
		
		House prices are mainly positively correlated with MF, typical areas include the southwest of Nanshan District, the southern coast of Baoan District, etc. Negatively correlated areas include Huanggang, in Futian District along the border with Hong Kong. We speculate that marginal districts may have a stronger positive correlation between management fees and house prices.
		
		\begin{figure}[H]
			\centering
			\includegraphics[width=0.5\textwidth]{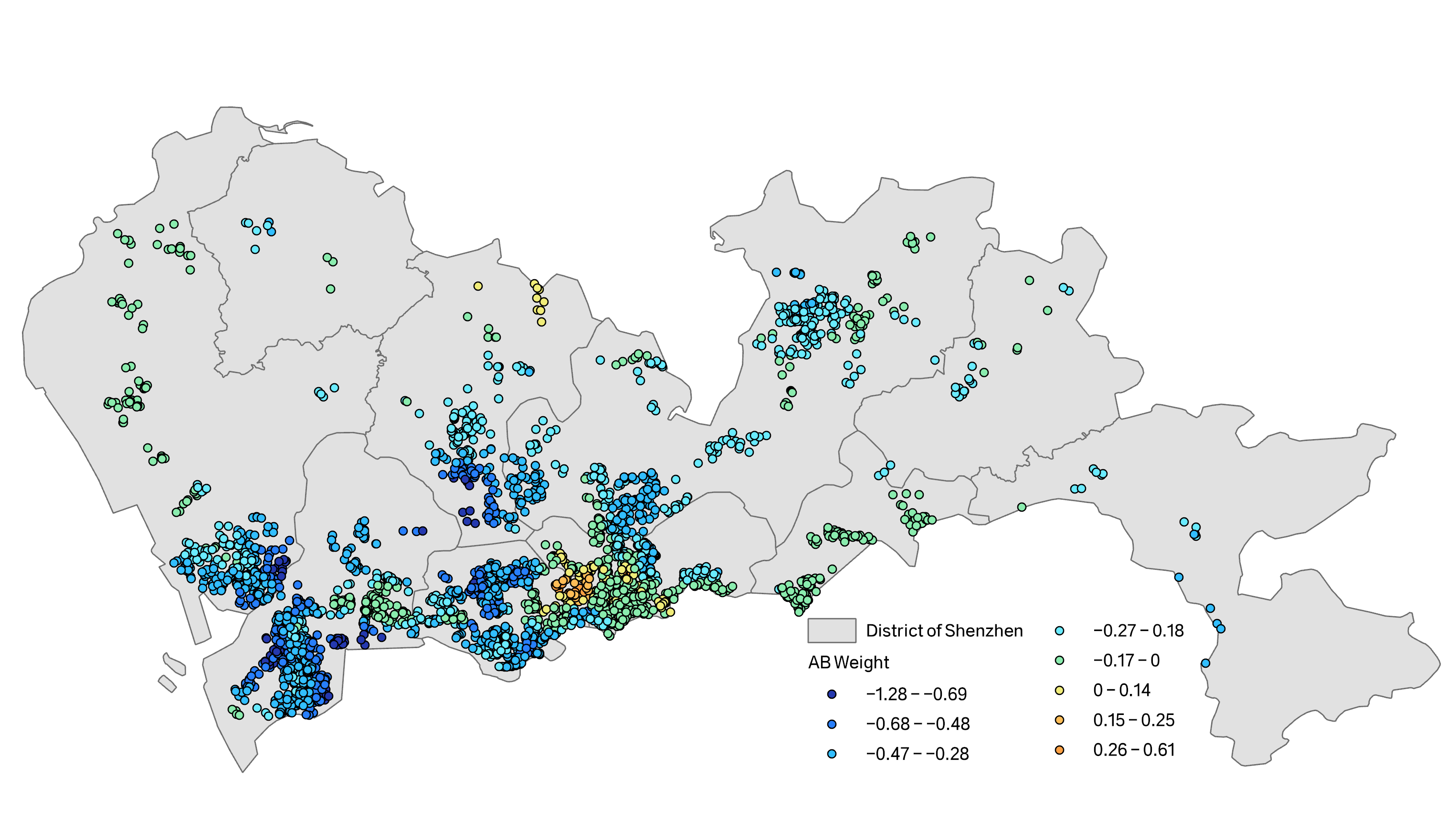}
			\caption{AB Weight Distribution}
		\end{figure}
		
		The growth of AB has had a restraining effect on house prices in Shenzhen for the most part. The negative correlation between house prices and AB is strongest in the coastal Nanshan District with Houhai as the core, the central Futian District with Xiangmi Lake's eastern shore as the core, and the southern Longhua District with Shenzhen North Station as the core. This may be due to the large supply of quality new houses near these locations, and the relatively old properties are vulnerable to the cold market. At the border of Luohu and Futian districts, the effect of AB on house prices shifts from a negative to a rare positive correlation, i.e. properties here do not have discounts due to old age, but may instead have premiums. According to the research of Goodman et al., the process by which house age affects house prices is nonlinear, with a positive effect on house prices when the age of the house is greater than a certain threshold.\cite{ref37} In fact, this area explored by the GNNWR model is exactly the area where the earliest construction in Shenzhen took place, and the famous landmarks Dongmen Old Street and Diwang Building are located near this area.
		
		\begin{figure}[H]
			\centering
			\includegraphics[width=0.5\textwidth]{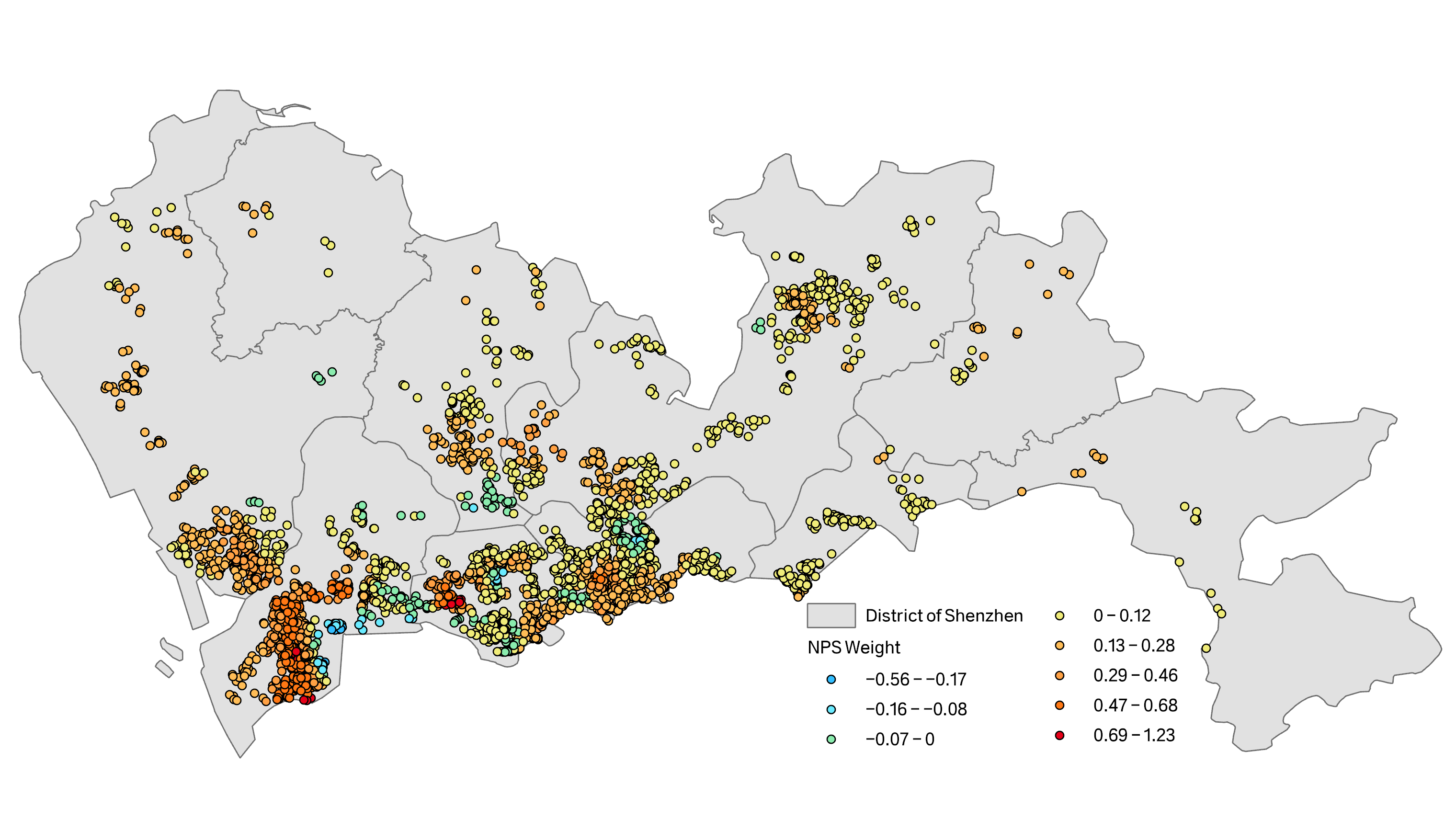}
			\caption{NPS Weight Distribution}
		\end{figure}
		
		NPS and house price are basically positively correlated. The strongest positive correlations are found in the central Nanshan District, Xiangmi Park in western Futian District, and near Caiwuwei in Luohu District. It can be speculated that the contribution of NPS to house prices should be most obvious in middle-class residential areas and wealthy areas. The high positive correlation areas explored by the GNNWR model are consistent with the distribution of middle-class residential areas and wealthy areas.
		
		\begin{figure}[H]
			\centering
			\includegraphics[width=0.5\textwidth]{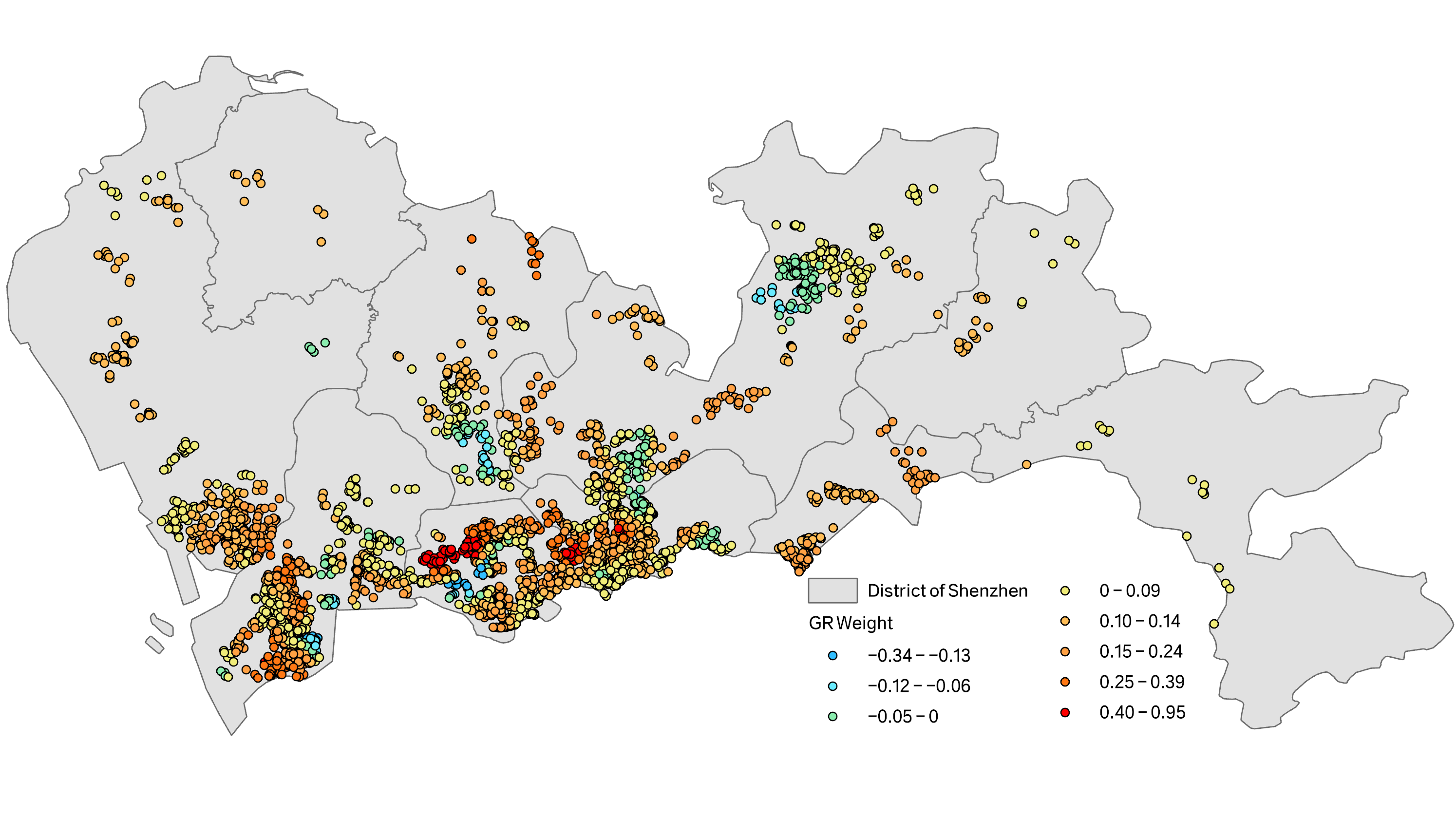}
			\caption{GR Weight Distribution}
		\end{figure}
		
		The increase of GR can raise the house price, especially for the central Futian District and the central Luohu District, which are located in the prosperous part of the city with higher demand for GR. In the suburbs and along the coast, GR has a smaller effect on raising house prices, and there is even a subtle negative correlation zone in Longgang District.
		
		\begin{figure}[H]
			\centering
			\includegraphics[width=0.5\textwidth]{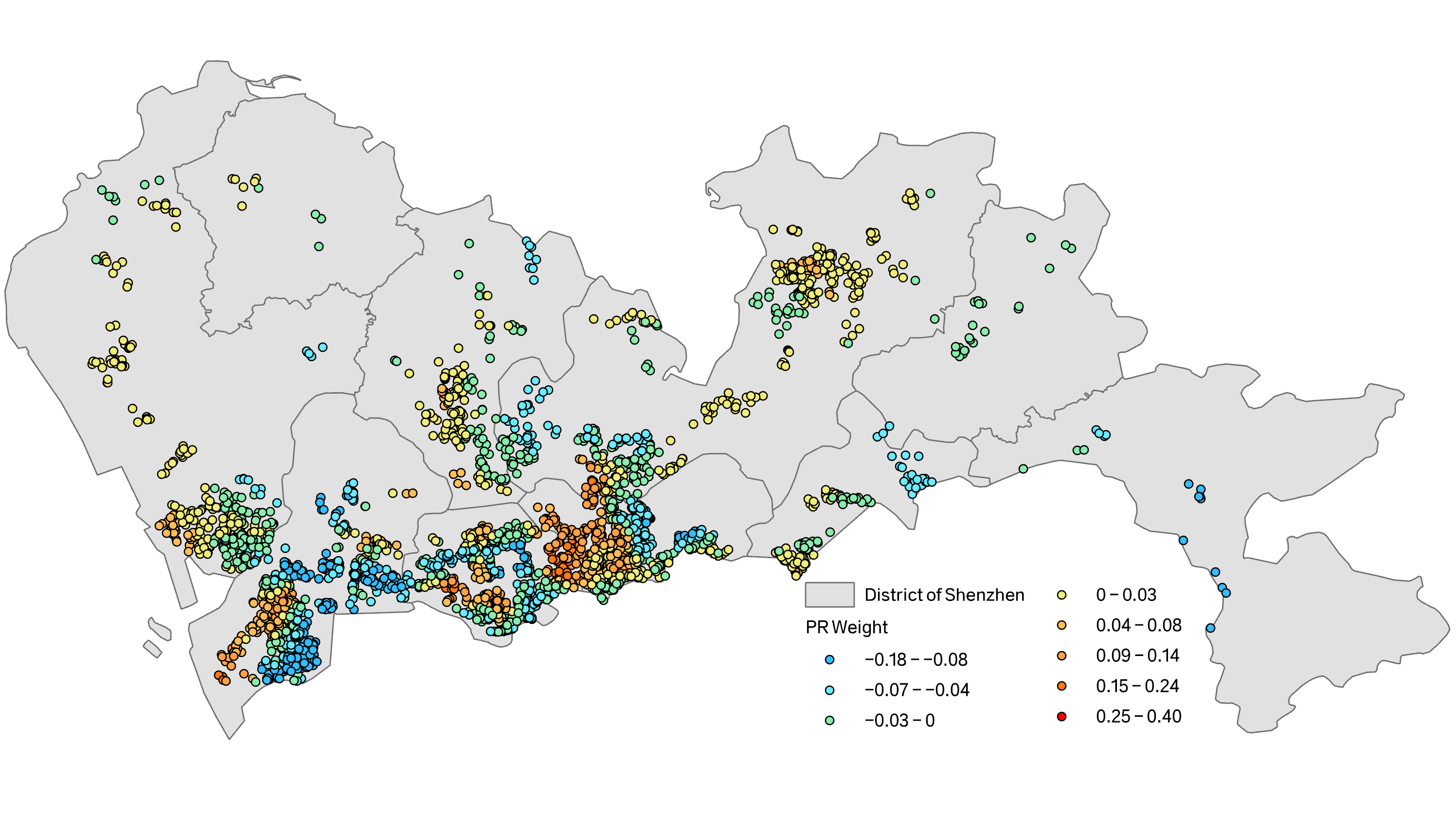}
			\caption{PR Weight Distribution}
		\end{figure}
		
		The fluctuation of PR is relatively small, and its impact on the house price is not significant from the sight of average weight. However, there is an area with clear positive correlation between PR and house price, the western part of Luohu District. This is contrary to the general perception, and we believe that it is mainly because the plot ratio there is closely related to the overall appearance of the neighborhood. The western part of Luohu District is the older urban area of Shenzhen, and a low plot ratio tends to represent the old and dilapidated character of the neighborhood, while a high plot ratio tends to be able to correspond to new high-rise housing. Probably for this reason, a positive correlation area appears here, while in other locations it does not.
		
		\subsubsection{Analysis of Environment related variables}
		The environmental variables considered in this model include SD, DSS, NSS, QASP.
		
		\begin{figure}[H]
			\centering
			\includegraphics[width=0.5\textwidth]{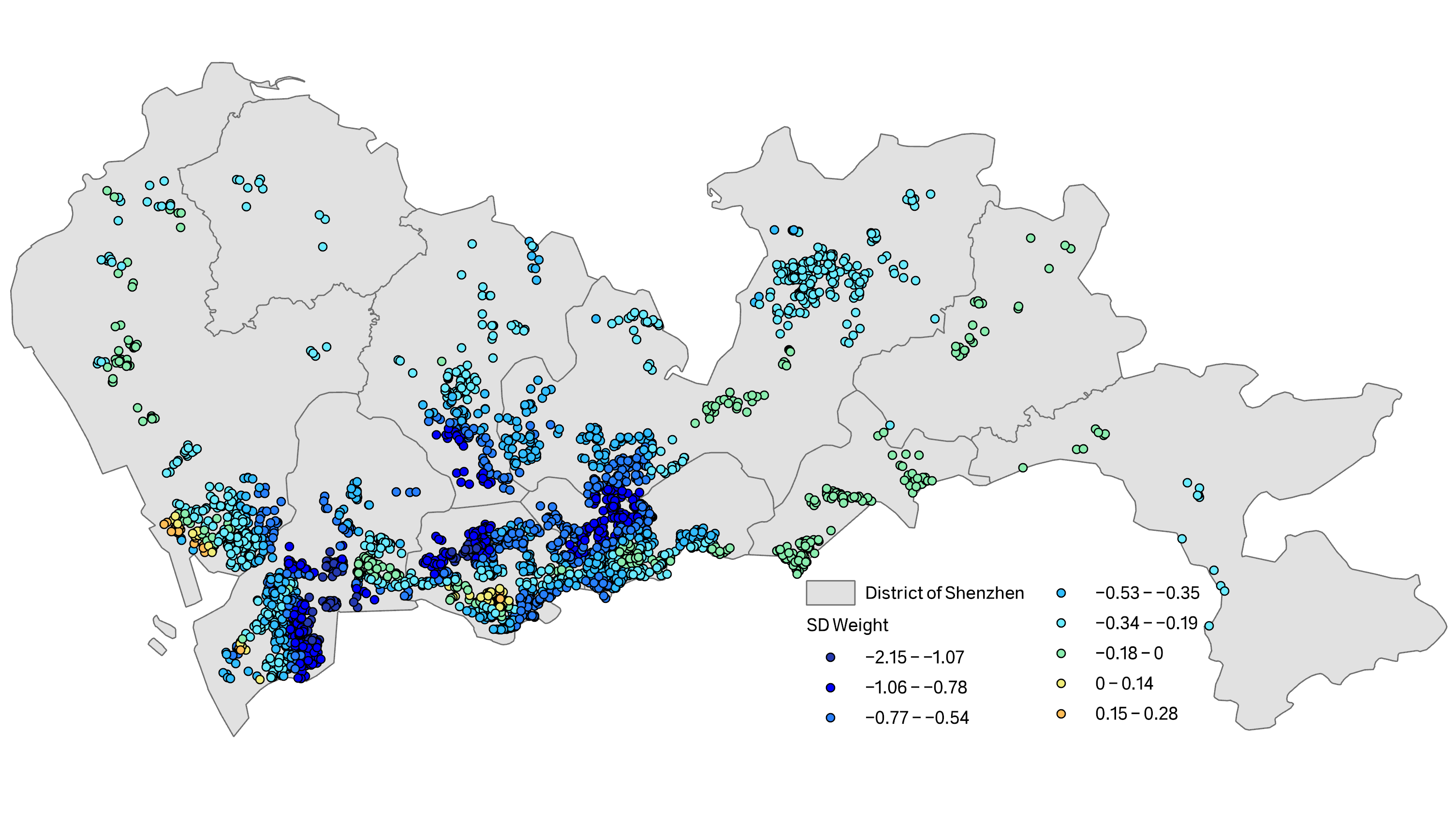}
			\caption{SD Weight Distribution}
		\end{figure}
		
		There is a very obvious negative correlation between SD and house prices, and the most typical areas include most of Nanshan District, most of Luohu District, etc. In comparison, the negative impact of SD in suburban and inland areas is slightly smaller, including Pingshan District and Guangming District. We speculate that suburban areas farther from the sea have other natural landscape, such as lakes and forests, which partially compensate for the disadvantage of being farther from the sea. Also, here SD is already quite large, and the absolute value of the coefficient need not be large to fully reflect the weakening effect on house prices. It should be noted in particular that in certain inland parcels, there is also a prominent negative SD correlation, such as the southern part of Longgang District and the northern edge of Luohu District. We believe that this is due to the fact that SD here actually characterizes the distance from the core urban area, thus triggering a strong negative correlation.
		
		\begin{figure}[H]
			\centering
			\includegraphics[width=0.5\textwidth]{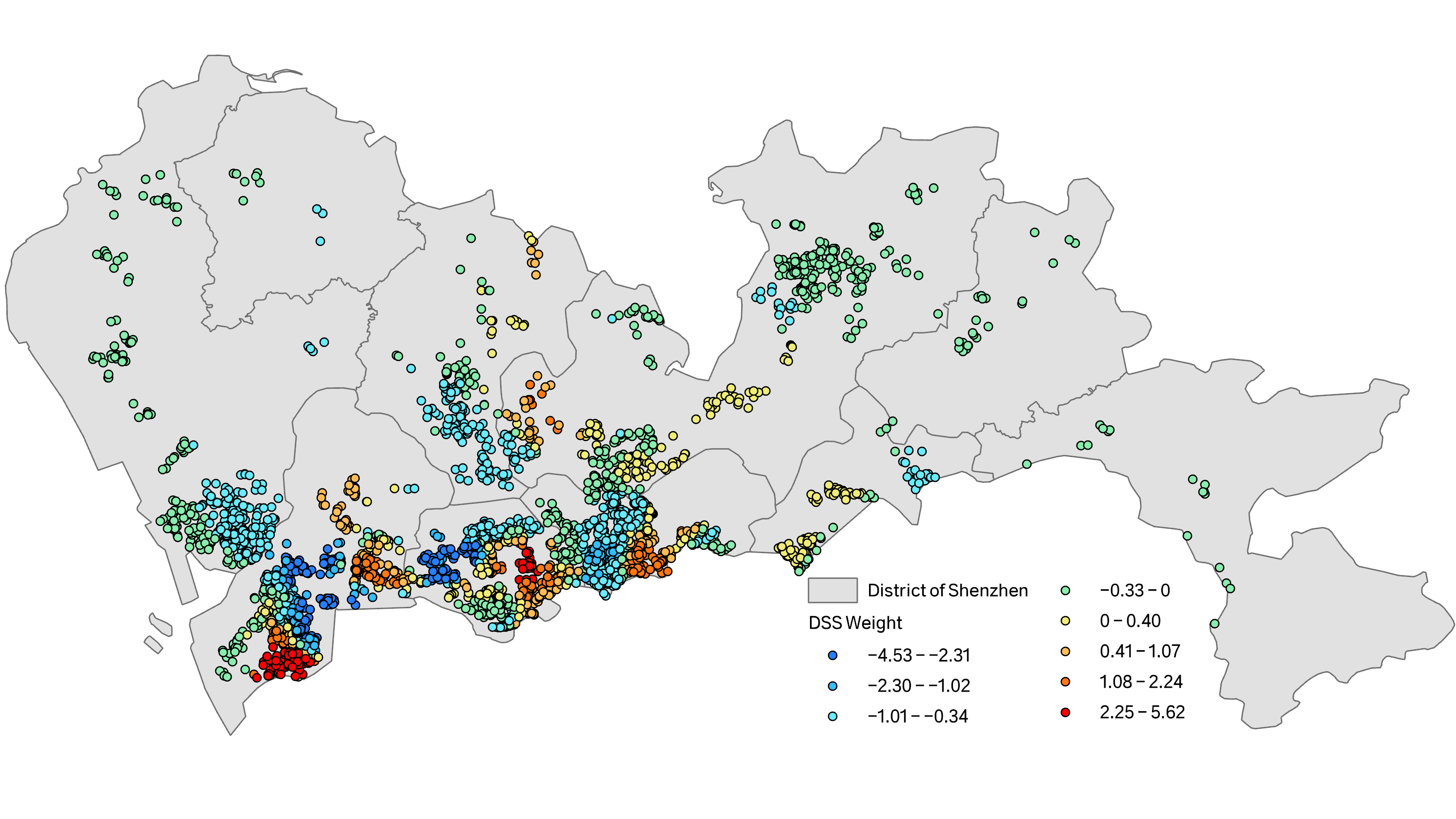}
			\caption{DSS Weight Distribution}
		\end{figure}
		
		The correlation between DSS and house price fluctuates very much. Generally, further away from the subway means lower house price. The regions showing negative correlation include most of the suburbs, especially the southern part of Baoan District and the southern part of Longhua District. In the main urban area, the western part of Luohu District, the central part of Futian District, and the central part of Nanshan District have significant negative correlation between house price and DSS.
		
		However, the areas where DSS is positively correlated are the southeastern part of Futian District, the southern part of Luohu District, and the southern part of Nanshan District. After inspection, a large number of jobs are concentrated in these areas. Southeastern part of Futian District corresponds to the Huaqiang North Market, one of the biggest Cell phone parts distribution markets around the world. The southern part of Luohu District corresponds to Xinxiu Village Industrial Zone.  The southern part of Nanshan District corresponds to the area around Shekou Industrial Zone.
		
		From this we have the following inference. On the one hand, for somewhere like residential areas and suburban areas, the closer to the subway entrance, the more convenient the commuting will be, and the price of housing will naturally have a certain increase. On the other hand, directly above the subway entrance, too much movement of people and underground vibration of the subway may have a negative impact on the price of housing. Moreover, for areas with dense subway entrances, CBD or industrial areas, being too close to the subway entrances may have a negative impact on house prices. People who buy houses in this neighborhood are already close to their workplace, and the need to commute with the help of the subway is insignificant; being too close to the subway entrance will instead aggravate the noise and congestion.
		
		\begin{figure}[H]
			\centering
			\includegraphics[width=0.5\textwidth]{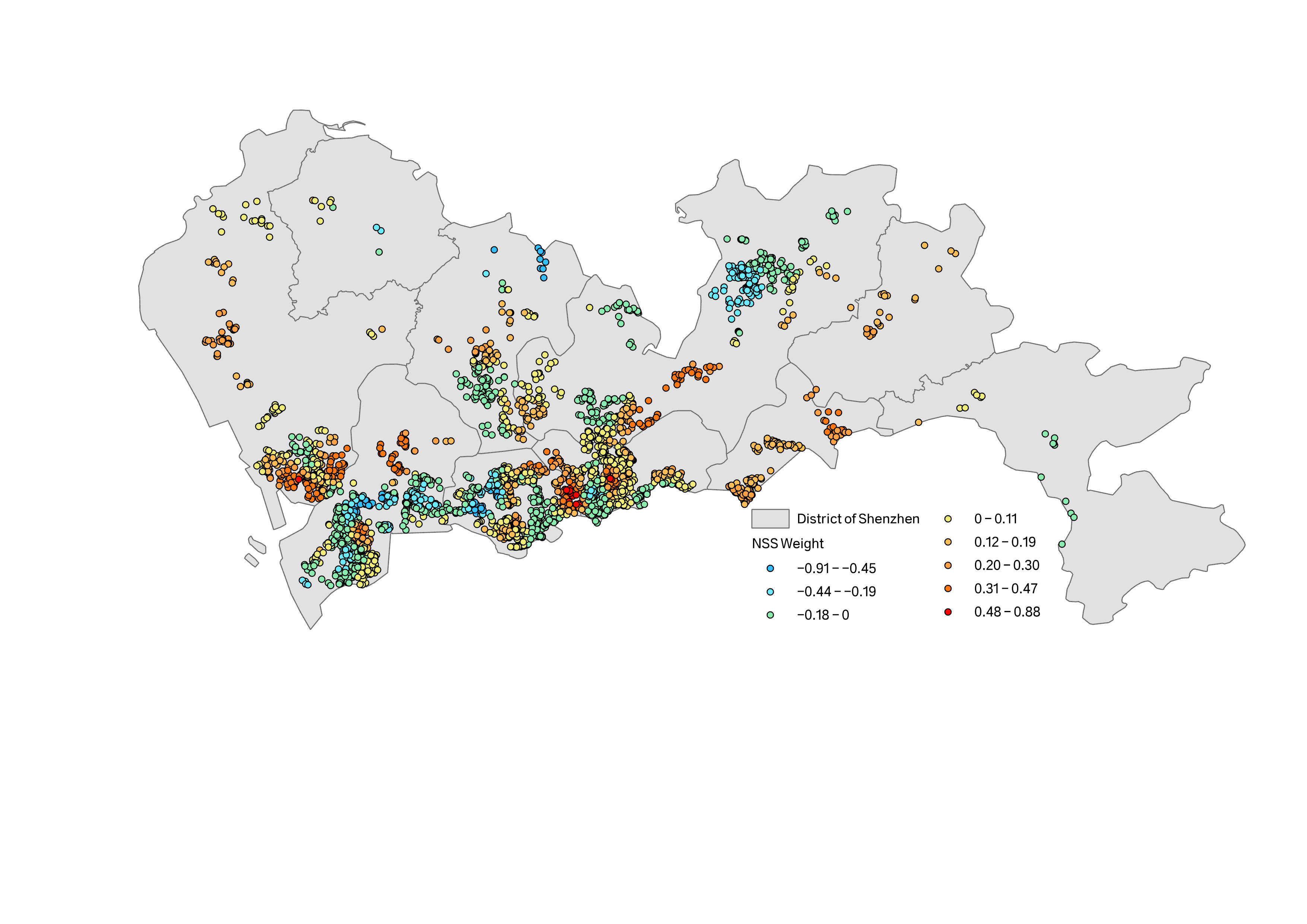}
			\caption{NSS Weight Distribution}
		\end{figure}
		
		The relationship between NSS and house price is that the more subway entrances there are, the higher the house price will be. Specifically for each district, we can find that the distribution of positive and negative correlations is almost opposite to that of DSS in Nanshan District, Futian District and Luohu District. This result confirms our above conjecture.
		
		\begin{figure}[H]
			\centering
			\includegraphics[width=0.5\textwidth]{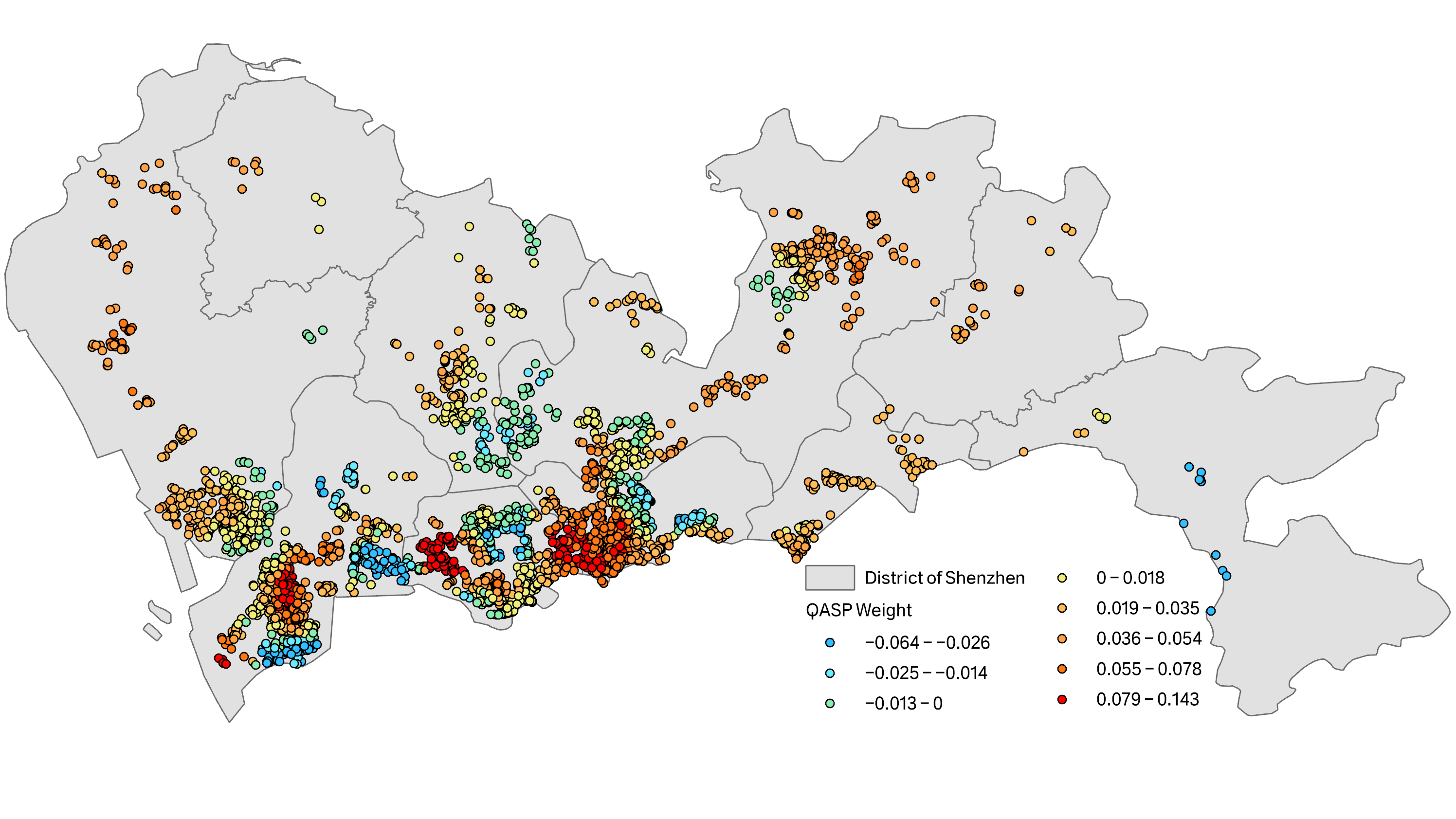}
			\caption{QASP Weight Distribution}
		\end{figure}
		
		Surprisingly, the relationship between QASP and house prices is relatively weak. Except for the western part of Luohu District, where house prices are strongly positively correlated with QASP, the effect of QASP on house prices is not significant in all other districts in Shenzhen. We believe this is due to the fact that the western part of Luohu District is the old city of Shenzhen, which makes many old residential areas sell themselves by highlighting its mature school districts, resulting in a strong positive correlation effect. In contrast, the new district's high-quality new housing does not have an established school district, and other factors dominate, the impact of the school district is relatively weaker.

		\section{Conclusion and Discussion}
		In the study based on Shenzhen house price data, we have ample evidence to prove the superiority of the GNNWR model over the OLS and GWR models. We use a ten-fold validation approach, and the following results are obtained by predicting for one-tenth of the data each time. Using RMSE as the standard, GNNWR improved by 13\% compared to GWR and 47\% compared to OLS. In terms of all other indicators, the GNNWR model shows significant improvement compared to both GWR and OLS. Second, we also performed sufficient tests to demonstrate the robustness of the GNNWR model. The mechanism of ten-fold validation avoids stochastic interference, and tests performed on the test set further demonstrate that the model is fully valid. In the section on hypothesis testing, we analyzed the significance of spatial heterogeneity. Compared with the GWR model, which also models spatial heterogeneity, and judged with the help of the AICc metric of the training set, it can be found that the improvement in fitting accuracy of the GNNWR model compared with the GWR model is much greater than the improvement in the complexity of the spatial weight matrix. All of these analyses clearly show that GNNWR has good robustness. Finally, we also analyze the spatial heterogeneity explored by GNNWR, which corroborates the outstanding information mining ability of GNNWR model in the context of Shenzhen.
		
		This study focuses on the following innovations. First, GWR, as a relatively traditional modeling method for spatial analysis, commonly used kernel functions only have two choices of bi-square and Gauss. Therefore, the calculated spatial weight matrix often does not adequately reflect the dataset characteristics, which is the original intention of GNNWR being proposed. Second, currently, other studies on modeling house prices with the help of neural networks, hardly introduce a ten-fold validation mechanism. This is a serious problem, and this study was refined based on more mature experimental specifications for neural networks. Third, as some scholars have suggested, the "black box" approach of neural networks has significantly limited the practical significance of neural networks in predicting house prices.\cite{ref38} Both polynomial regression models and traditional neural network methods depart from the linear structure and have relatively complex expressions, making the analysis and prediction much more difficult. Fourth, neural network prediction methods, that take less geographical location information into account, make their performance unstable. A part of the study highlighted the accuracy of neural network prediction compared to OLS models,\cite{ref39,ref40,ref41} but some studies concluded that neural network models often fail to significantly outperform OLS model and its improved models (including hedonic models that correct the dependent variable by log and polynomial regression model).\cite{ref42, ref43} But in any case, the RMSE-based metrics show that there are few neural network models with more than 30\% improvement compared to OLS.
		
		Since the GNNWR model was proposed, there is no relevant applied research in the socioeconomic field, and this study fills this blank. Future improvements can be made in the following directions. First, the error term in the linear model can be further tested for the linearity, homoscedasticity, independence and normality properties. If they are not satisfied, the dependent variable can be pretreated using the Box-Cox method. Second, more independent variables can be obtained to further expand the choice of independent variables. Third, the independent variables can be preprocessed and filtered. For example, if three independent variables, total number of buildings, total number of apartments, and total number of units, are obtained for a certain residential quarter, these independent variables will have multicollinearity. Using PCA, the principal components of these independent variables can be extracted and the multicollinearity can be reduced. Another example is to use more data, such as enrollment rate, distance to school, to evaluate a school district. The independent variables can also be filtered using the forward method, backward method or stepwise method. Fourth, comparable tests can be further performed on other data sets or data from multiple cities can be collected to build a house price prediction benchmark. Fifth, time series analysis can be added to make the GNNWR model have the function of prediction in both time and space.

	\bibliographystyle{acm}
	\bibliography{test}
	
	\end{multicols}
\end{document}